%% file: main_jctc.tex
\documentclass[journal=jctcce, manuscript=article]{achemso}

\usepackage[version=3]{mhchem} 

\input{notations}

\author{Xiaoxu Li}
\affiliation[BNUZ]
{Faculty of Arts and Sciences, Beijing Normal University, Zhuhai 519085, Guangdong, China.}
\author{Ge Xu}
\affiliation[BNU]{School of Mathematical Sciences, Beijing Normal University, Beijing 100875, China.}
\author{Huajie Chen}
\email{chen.huajie@bnu.edu.cn}
\affiliation[BNU]
{School of Mathematical Sciences, Beijing Normal University, Beijing 100875, China.}
\author{Xingyu Gao}
\email{gao\_xingyu@iapcm.ac.cn}
\affiliation[IAPCM]
{Laboratory of Computational Physics, Institute of Applied Physics and Computational Mathematics, Beijing 100088, China.}
\author{Haifeng Song}
\email{song\_haifeng@iapcm.ac.cn}
\affiliation[IAPCM]
{Laboratory of Computational Physics, Institute of Applied Physics and Computational Mathematics, Beijing 100088, China.}

\title[ML-MCTS for Configuration Generation]
{A Multi-Level Monte Carlo Tree Search Method for Configuration Generation in Crystalline Systems}

\begin{document}



\begin{abstract}
In this paper, we study the construction of structural models for the description of substitutional defects in crystalline materials. 
Predicting and designing the atomic structures in such systems is highly challenging due to the combinatorial growth of atomic arrangements and the ruggedness of the associated landscape.
We develop a multi-level Monte Carlo tree search algorithm to generate the ``optimal" configuration within a supercell.
Our method explores the configuration space with a expanding search tree through random sampling, which further incorporates a hierarchical decomposition of the crystalline structure to accelerate exploration and reduce redundancy.
We perform numerical experiments on some typical crystalline systems to demonstrate the efficiency of our method in identifying optimal configurations.
\end{abstract}

\section{Introduction}
\label{sec:introduction}

Generating the atomic configuration for complex materials system, such as doped semiconductors, solid solutions and multicomponent alloys, is a fundamental step in simulating their physical and chemical properties \cite{okhotnikov2016supercell, banik2023continuous,shinde2011direct,li2023machine}.
By assigning different atomic species to specific lattice sites in a supercell \cite{nieminen2006supercell}, one can model the material's structure and perform subsequent electronic structure calculations to predict its properties.
Despite its conceptual simplicity, the search for physically meaningful atomic configurations within the supercell remains highly challenging. 
The number of possible configurations grows combinatorially with the number of atomic sites and species, making exhaustive enumeration infeasible for realistic systems.
In practice, the desired configuration is generally obtained by minimizing an energy- or property-based objective function. However, this function often exhibits a complex and rugged landscape, posing a significant challenge for global optimization.

Standard approaches for tackling combinatorial optimization in configuration search can be broadly categorized into three types: the exact algorithms, the enumeration-based methods, and the stochastic methods.
Exact algorithms such as branch-and-bound \cite{morrison2016branch} and cutting plane methods \cite{khachiyan1980polynomial} offer theoretical guarantees of optimality but are computationally intractable for large-scale systems due to the exponential growth of the configuration space.
Enumeration-based methods generate all possible configurations within a supercell while applying symmetry reduction to eliminate redundant cases. This strategy is effective for small systems and is supported by tools such as the Supercell program \cite{okhotnikov2016supercell}, but it quickly becomes infeasible as the system size or number of atomic species increases.
The stochastic methods are more widely used in practice for more complex problems. 
The greedy methods iteratively select lower-energy configurations but are easily trapped in local minima. 
Basin-hopping techniques \cite{wales1997global} combine local minimization with random perturbations to escape shallow basins but may still struggle with complex landscapes. 
The Monte Carlo methods, including Metropolis-Hastings \cite{binder1992monte, landau2021guide} and simulated annealing \cite{kirkpatrick1983optimization}, use probabilistic acceptance rules to traverse the energy landscape, though their performance depends heavily on parameter tuning and cooling schedules.
The genetic algorithms \cite{katoch2021review, mccall2005genetic} evolve multiple candidates through crossover and mutation, but may quickly get stuck in suboptimal regions due to loss of diversity.
Recently, the machine learning–assisted heuristics \cite{bengio2021machine, mirshekarian2018machine} attempt to guide the search process adaptively, but they often require significant training data or model calibration, which can be costly or problem-specific.
Despite their scalability and empirical success, these methods generally lack global guarantees and may converge slowly or inconsistently when faced with high-dimensional, rugged energy landscapes.

In this work, we present a novel framework for configuration search in crystalline systems.
In particular, a multi-level Monte Carlo Tree Search (ML-MCTS) method is proposed to identify the optimal atomic arrangement in a given supercell with fixed composition. 
The Monte Carlo Tree Search method \cite{browne2012, swiechowski2023monte} was originally introduced in the field of game-playing, where it was successfully applied to board games such as Go \cite{fu2016alphago}. 
It has been extended to a wide range of applications such as scheduling and combinatorial optimization \cite{browne2012, swiechowski2023monte},
and has recently gained attention in materials science with applications in atomistic sampling and structure prediction \cite{banik2021learning, banik2023continuous, loeffler2021reinforcement}.
We develop a multi-level extension of the MCTS method, by incorporating a hierarchical decomposition of the action space. 
This hierarchical strategy enables the algorithm to prioritize large-scale atomic rearrangements and progressively refine configurations through localized adjustments, which significantly reduces redundant exploration and improves the convergence of optimization process.
We demonstrate the efficiency of the proposed method through numerical experiments on typical two-dimensional materials. 

\vskip 0.5cm

{\bf Outline.} 
The rest of the paper is organized as follows.
In Section \ref{sec:model}, we formulate the generation of supercell configuration as a combinatorial optimization problem.  
In Section \ref{sec:graph}, we provide a graph-based representation for the configuration optimization problem, particularly for the structure of configuration space and the idea of standard search algorithms.
In Section \ref{sec:mcts}, we develop a MCTS framework for searching the optimal configuration.
In Section \ref{sec:MLMCTS}, we propose a multi-level MCTS approach, which enhances  the search efficiency by introducing a hierarchical decomposition of the action space.
In Section \ref{sec:numerics}, we present some numerical experiments on two-dimensional hexagonal lattice systems to validate the efficiency of our algorithm. 
Finally, we give some conclusions in Section \ref{sec:conclusion}.

\section{Supercell configuration optimization}
\label{sec:model}
\setcounter{equation}{0}

We consider a crystalline system embedded in a $d$-dimensional ($d=1,2,3$) Bravais lattice $\LL = A\mathbb{Z}^d$, with a non-singular matrix $A \in \mathbb{R}^{d \times d}$.
The system consists of $s\in\N$ distinct atomic species, represented by the set $\X_s := \{X_1, \cdots, X_s\}$.
These atoms are distributed over a finite supercell $\Lambda=A[-D, D]^d\cap\LL$, with periodic boundary conditions imposed to replicate the configuration in all spatial directions.
We denote by $N:=\#\Lambda$ the total number of sites in this supercell, and by $N_k$ the number sites for atomic species $X_k$ with $k=1,\dots,s$.
We then define the admissible set for possible configurations by
\begin{eqnarray}
\mathbb{S} = \bigg\{
\config \in \{X_1,\cdots,X_s\}^{\Lambda}~:~\#\{\ell\in\Lambda:\sigma_{\ell}=X_k\} = N_k \quad\text{for}~k=1,\cdots,s 
\bigg\}.
\end{eqnarray}
For instance, when $s=2$ a configuration corresponds to a discrete assignment of two atomic species to the lattice sites, such that exactly $N_1$ sites are occupied by $X_1$, and the remaining $N_2=N-N_1$ sites are occupied by $X_2$.

To identify the ``optimal” configuration, we need an objective function $f: \mathbb{S} \to \R$ associated to some physical quantities of interest.
Then the optimal configuration is defined as a solution to the following optimization problem:
\begin{eqnarray}
\label{model-optimization-eq}
\min_{\config\in\mathbb{S}} f(\config).
\end{eqnarray}
Note that the desired optimal configuration is determined by the objective model.
If the goal is to find an equilibrium state, one seeks a configuration that minimizes the total energy, which could be computed by electronic structure models or empirical atomic potentials.
If the goal is to reproduce a statistical properties of a random alloy, one needs to generate a configuration that matches some target correlation functions, as in the special quasirandom structure (SQS) model \cite{van2013efficient} or the similar atomic environment (SAE) model  \cite{tian2020structural}.

We observe that \eqref{model-optimization-eq} is a typical combinatorial optimization problem, where the difficulties arise from at least two aspects:
(a) The size of $\mathbb{S}$ is given by the multinomial coefficient
\begin{equation}
\binom{N}{N_1} \binom{N-N_1}{N_2}\cdots \binom{N-\sum_{k=1}^{s-1} N_k}{N_s} = \frac{N!}{N_1! N_2!\cdots N_s!},
\end{equation}
which grows dramatically fast with respect to $N$; and
(b) The objective function $f$ generally exhibits a complex and rugged landscape.

\section{Graph representation}
\label{sec:graph}
\setcounter{equation}{0}

\begin{figure}[htb!]
	\centering
	\subfigure[Configuration graph]{
	\includegraphics[width=0.8\textwidth]{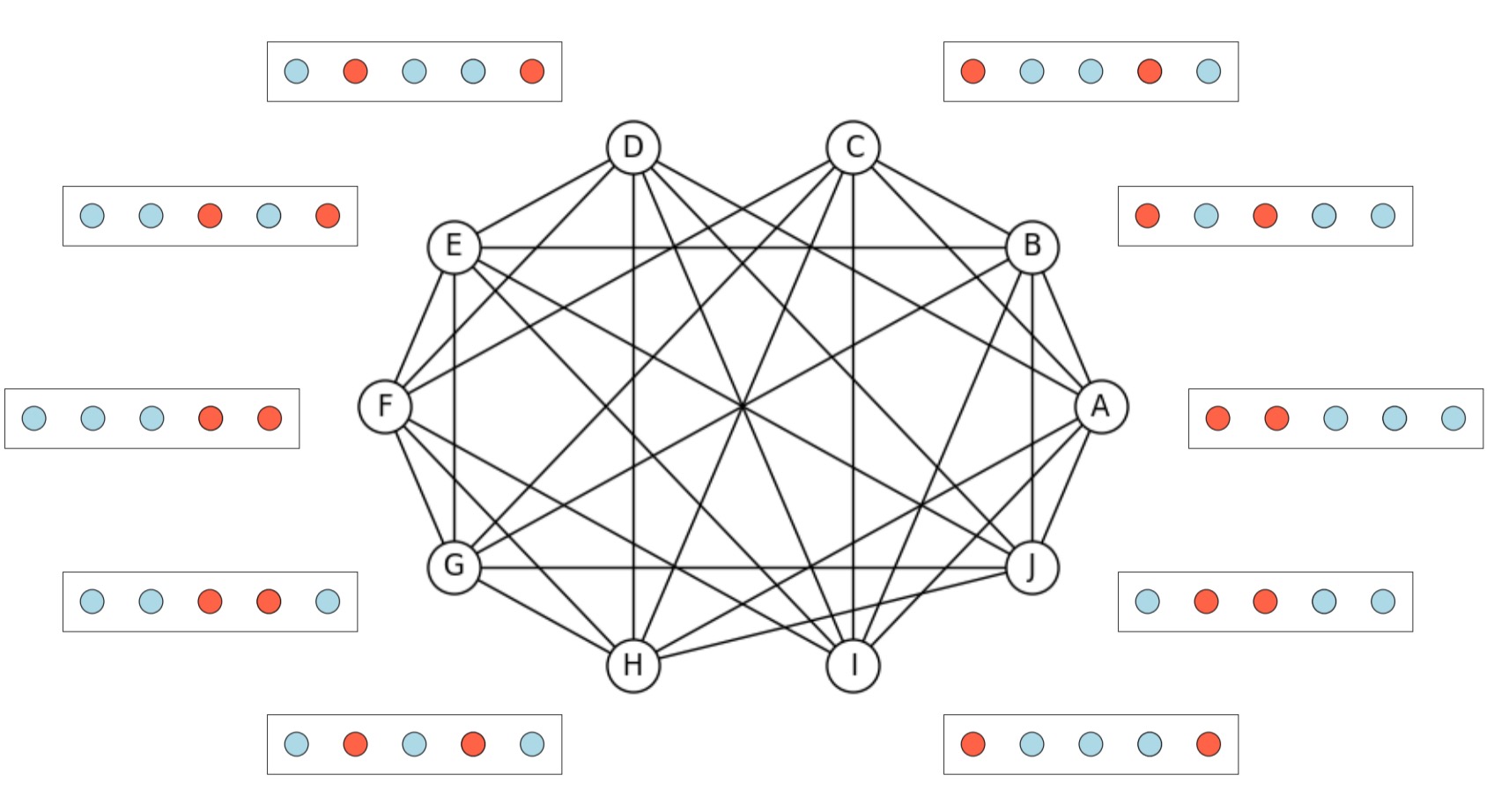}}
    \vskip 0.2cm
	\subfigure[Search path]{
    \includegraphics[width=0.8\textwidth]{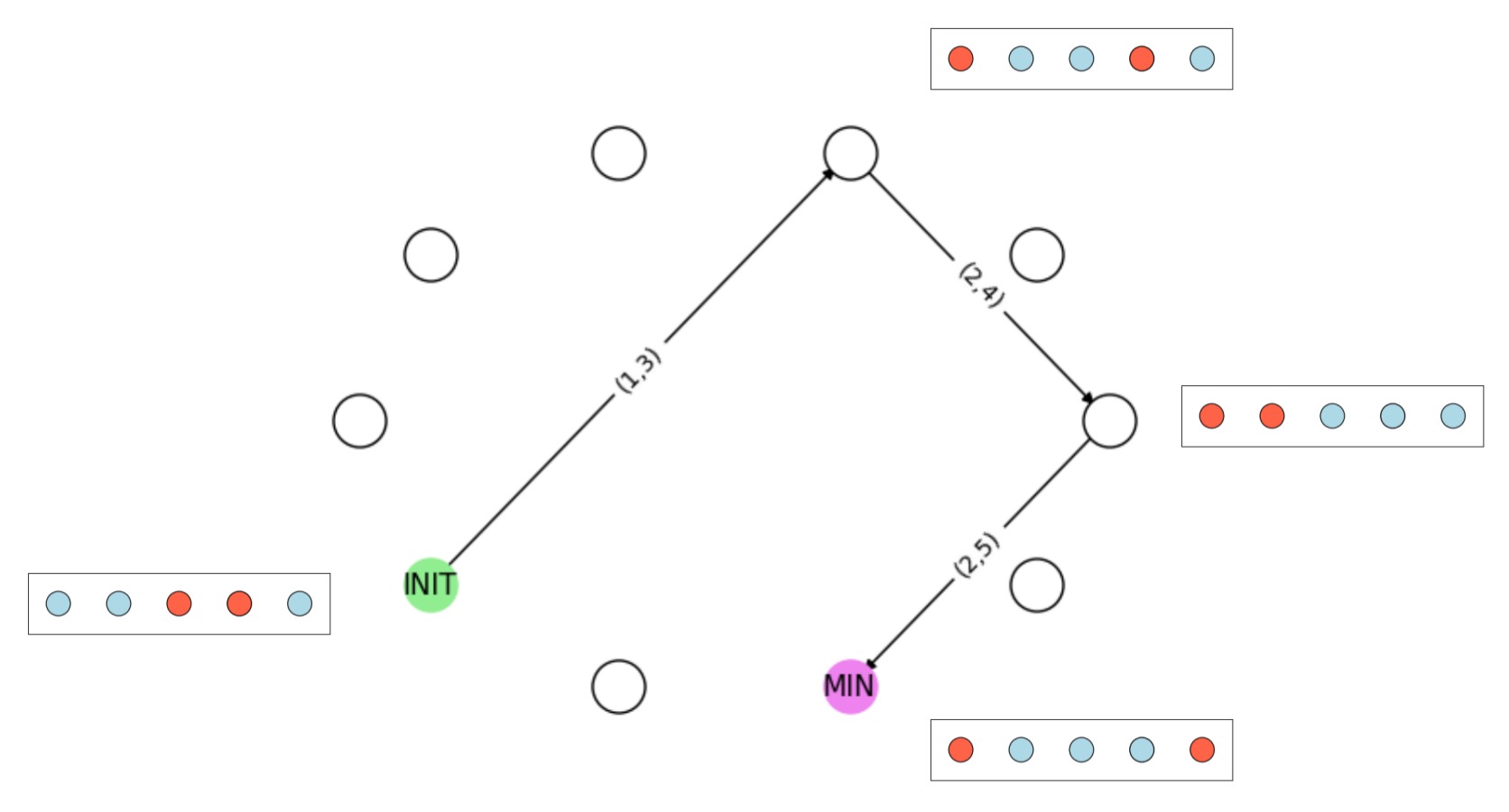}}
    \caption{Graph representation for an one-dimensional binary compound with $N=5$ and $N_1=2$, $N_2=3$.
    There are 10 different configurations and each one has 6 adjacent vertices.
    (a) The graph with all vertices and edges. (b) A search path from an initial vertex to the optimal one.
    }
    \label{fig:graph}
\end{figure}

We will formulate the combinatorial optimization problem \eqref{model-optimization-eq} and the associated algorithms within a graph representation.
With a given supercell $\Lambda$, each configuration $\config \in \mathbb{S}$ is represented as a vertex in a graph. 
Two configurations are considered adjacent and connected by an edge if they differ by exactly one pair of site occupations.
Each edge also corresponds to an {\it action} that transforms one configuration into another by swapping the occupations of two sites.
In particular, for a configuration $\config \in \Sset$, the set of all possible actions on $\config$ is given by the action space
\begin{equation}
\A(\config) = \Big\{(p,q) \in\Lambda\times\Lambda~:~\sigma_p\neq\sigma_q \Big\} .
\end{equation}
We show the graph of an one-dimensional binary compound example with $N = 5$, $N_1=2$ and $N_2=3$ in Figure \ref{fig:graph}(a).
We remark that more general transitions involving multiple-site swaps can also be used while this work considers only the actions with single-pair exchange. 
Our goal is to find the the optimal configurations of \eqref{model-optimization-eq}, which correspond to vertices with the smallest $f$-value.
Since the number of vertices in the graph is prohibitively large in most cases, exhaustively enumerating all $f$-values is infeasible.

The general search strategy resembles navigating a maze, which starts from a random initial vertex and attempt to find a path leading to the vertex with the lowest $f$-value (see Figure \ref{fig:graph}(b) for a search path). 
The overall cost of the search algorithm is primarily determined by the number of objective function evaluations, which corresponds to the number of visited vertices during the search process, including the repeated visits.
An efficient search method should aim to reduce the number of evaluations as much as possible, while effectively guiding the search toward configurations with lower $f$-values.

With the graph representation, one can easily visualize how existing algorithms navigate through the configurations/vertices. 
The greedy algorithm simply finds the ``best" adjacent vertex that has the lowest $f$-value and move to it.
This approach often gets stuck in some local minima, whose all adjacent vertices have higher $f$-values, but there is a vertex in the graph away from it possessing a even lower value.
To escape local traps, the basin-hopping method alternates between greedy descent and random perturbations. 
After reaching a local minimum, this method swaps several distant site occupations to jump to a new region in the graph and restarts the greedy algorithm.
The Metropolis-Hastings (MH) method allows occasional uphill moves controlled by a temperature parameter, selecting a random adjacent vertex and move to it with a probability based on the change of $f$-value.
The simulated annealing method exploits a similar idea, but behaves more like the greedy algorithm along the search with a decreasing temperature. 

The above mentioned methods can be understood within a unified tree search framework, that explores the configurations in the graph by expanding a tree from an initial guess.
Although they apply different strategies to add new vertices to the search tree, all these methods tend to rely on local environments of the current configuration, i.e. the $f$-values of adjacent vertices. 
Therefore, these methods somehow lose the ability to see beyond the immediate neighborhood, which limits their effectiveness in complex landscapes. 

To overcome this shortcoming, we will resort to a more ``global" strategy for the tree search framework, which can exploit the $f$-values of distant vertices in the graph. 
Based on this idea, we develop a Monte Carlo Tree Search algorithm for \eqref{model-optimization-eq}, building a search tree that could explore the configuration space more efficiently and thoroughly.

\section{Configuration search with Monte Carlo tree search}
\label{sec:mcts}
\setcounter{equation}{0}
\setcounter{figure}{0}

In this section, we propose a Monte Carlo Tree Search (MCTS) algorithm to find the optimal configuration of \eqref{model-optimization-eq}. 
The central idea of the MCTS algorithm is to expand an asymmetric search tree, which concentrates on more promising subtrees that are likely to contain optimal solutions.

In the search tree, each node represents a configuration (vertex) in the graph, and each edge corresponds to an action connecting two adjacent configurations. 
Note that the same vertex in the configuration graph may appear multiple times in the tree, as different action sequences can lead to the same configuration.

There are several important properties for the nodes, which are crucial for the MCTS algorithm.
First, each node $\config$ in the tree could have a number of child nodes.
The number of children determines the width of the tree, for which a maximum number $n_{\rm child}\in\N$ is set in the algorithm.
If a node in the search tree has fewer than $n_{\rm child}$ children, then this type of nodes is called expandable. 
Starting from the root, the algorithm recursively selects one child at a time until it reaches such an expandable node.
Second, each node $\config$ in the tree is associated with a value function $Q(\config)$, with $Q:\Lambda\rightarrow\R$ indicating how promising this node is towards the optimal configuration.
It could be initialized by $Q(\config) = -f(\config)$ when the node is added to the tree and is updated during the iterations as more information becomes available.
Finally, we will also assign an exploration function to the node in the search tree. 
For a node $\config$ in the search tree, let $\config_{\rm p}$ be its parent node and
\begin{equation}
\label{def:exploration}
R(\config_{\rm p}, \config) := C_{\rm U}  \sqrt{\frac{ \ln v(\config_{\rm p})}{v(\config)} },
\end{equation}
where $v:\Lambda\rightarrow\N$ counts the number of times this node have been ``visited", and $C_{\rm U}>0$ is a pre-set parameter for the trade-off between exploration and exploitation.

\begin{figure}[htb!]
	\centering
	\includegraphics[width=15.cm]{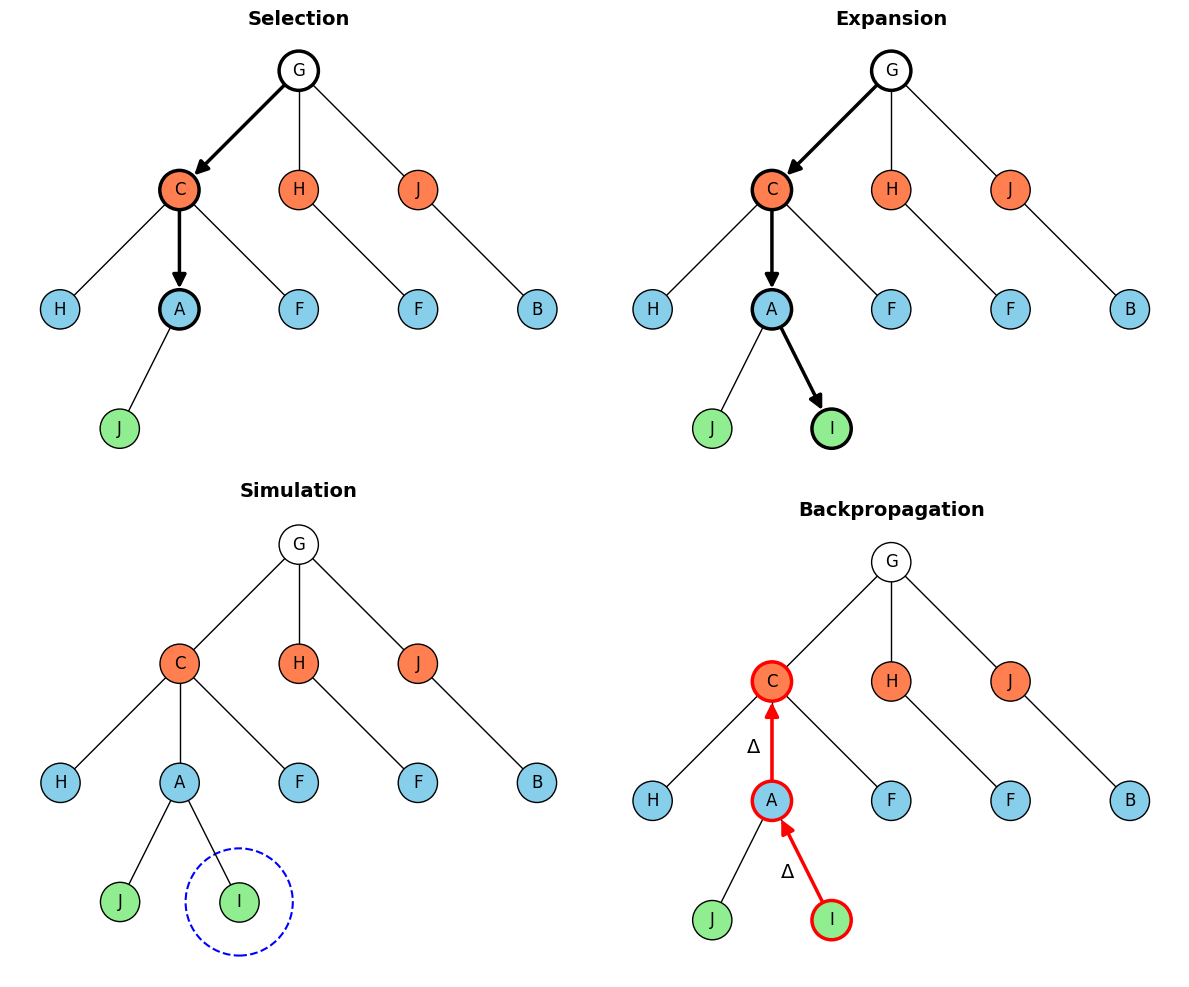}
	\caption{Illustration of the four stages in the MCTS algorithm.
    The letters on the nodes correspond to configurations in Figure \ref{fig:graph}(a). 
    The maximum number of child nodes per node $n_{\rm child}$ is set to 3. 
    Black arrows indicate the selection and expansion directions, while red arrows represent the backpropagation updates. 
    The dashed circle marks the node where the objective function $f$ is evaluated during the simulation step.}
	\label{fig:MCTS}
\end{figure}

The MCTS algorithm starts from an initial configuration as the root node, and gradually expands the tree by using some search policy.
In each iteration step, the algorithm proceeds through the following four stages, see also Figure \ref{fig:MCTS}. 
\begin{enumerate}
\item
Selection: From the root node, recursively chooses a child node $\config_{\rm c}$ that maximizes the following score with respect to its parent $\config_{\rm p}$
\begin{equation}
\label{UCB-def}
Q(\config_{\rm c}) + R(\config_{\rm p}, \config_{\rm c}),
\end{equation}
and obtain an expandable node $\config$ of the current tree.
\item 
Expansion:
Add a new node to the search tree by selecting a (random) configuration $\config_{\rm new}$ that is adjacent to $\config$.
\item 
Simulation:
Evaluate the new configuration, with the $f$-value denoted by $\Delta:=f(\config_{\rm new})$.
\item 
Backpropagation:
Pass the simulation result back along the path chosen during the earlier selection stage. 
For each node $\tilde{\config}$ on the path, its $Q$-value is updated by
\begin{eqnarray}
\label{update:Q}
Q(\tilde{\config}) \leftarrow \max\big\{Q(\tilde{\config}), -\Delta\big\}.
\end{eqnarray}
\end{enumerate}

In the selection stage, the first term in \eqref{UCB-def} represents the estimated quality of the child node at the current iteration; the second term promotes exploration by assigning higher scores to less frequently visited nodes. 
These two terms together guide the selection process by balancing the exploitation of known good configurations and the exploration of new, potentially better ones.
In the backpropagation stage, the update \eqref{update:Q} increases the $Q$-values of the nodes along the path if the new configuration has a low $f$-value. 
As a result, these nodes are more likely to be chosen in future iterations, helping guide the search toward more promising branches.

In practical implementations, MCTS starts with an empty tree that contains only the root node. 
In each iteration step, a new node is added to the tree through the above four-stage procedure. 
The algorithm runs under a computational budget, either by a maximum number of iteration steps or a maximum running time limit. 
The search stops when the budget is fully used, and the node with the lowest $f$-value throughout the tree is chosen as the best configuration. 

There are two hyper-parameters that are crucial for the efficiency of the MCTS algorithm.
The first parameter is $n_{\rm child}$ that controls the width of the search tree.
A larger value of $n_{\rm child}$ enables broader exploration and increases the chance of escaping local minima. 
In contrast, a smaller value often leads to more efficient search in the early stages by reducing unnecessary branching.
The other hyper-parameter is the trade-off parameter $C_{\rm U}$ in \eqref{def:exploration}.
In the selection stage, a larger $C_{\rm U}$ favors less-visited nodes, promoting a more balanced tree expansion; while a smaller $C_{\rm U}$ encourages repeated visits to nodes with higher $Q$-values.
In our numerical experiments (see Section \ref{sec:numerics}), we also apply an adaptive strategy for tuning $C_{\rm U}$, based on the idea that emphasizes exploration in the early iterations and gradually shifts toward exploitation as the search progresses.

Compared to standard approaches such as greedy algorithm, basin-hopping, and the MH method, the MCTS algorithm is not limited to the $f$-values within a local neighborhood.
It offers a more global perspective, and therefore could significantly reduce the risk of getting trapped in local minima.
However, the search could be quite redundant during the iteration, since there could be multiple nodes in the tree representing the same configuration, even at the same level and along the same path.
Therefore, we will develop a multi-level strategy for the MCTS algorithm to resolve this issue.

\section{Multi-level MCTS approach}
\label{sec:MLMCTS}
\setcounter{equation}{0}
\setcounter{figure}{0}

In this section, we propose a multi-level MCTS (ML-MCTS) framework by exploiting a hierarchical decomposition of the crystalline structure.
At each level $\ell\in\N$, with a maximum level $L$ corresponding to the finest decomposition, we decompose the supercell $\Lambda$ into non-overlapping subsets
\begin{equation*}
\Lambda = \bigcup_{k}\Lambda^{(\ell)}_k 
\qquad \text{satisfying} \qquad \Lambda^{(\ell)}_{k_1} \cap \Lambda^{(\ell)}_{k_2} = \emptyset \quad \forall~k_1\neq k_2.   
\end{equation*}
Moreover, the decomposition is nested across levels. Each subset $\Lambda^{(\ell)}_k$ at level $\ell$ arises from the refinement of a subset $\Lambda^{(\ell-1)}_{k'}$ for some $k'$ at the previous level.
Given this structure, we define a restricted set of actions at each level. For any configuration $\config \in \mathbb{S}$, the admissible actions at level $\ell$ are limited to
\begin{equation*}
\label{action-l}
\A^{(\ell)}(\config) := \Big\{(p,q) \in \Lambda^{(\ell)}_{k_1}\times\Lambda^{(\ell)}_{k_2}:~\sigma_p\neq \sigma_q,~k_1\neq k_2,~{\rm and~both}~\Lambda^{(\ell)}_{k_1}~\text{and}~\Lambda^{(\ell)}_{k_2}~{\rm lie~in~same~} \Lambda^{(\ell-1)}_{k} \Big\}.
\end{equation*}

Based on this multi-level structure, we modify the standard MCTS algorithm by introducing level-dependent action restrictions. 
Each node in the search tree is assigned an index $\ell = (h \bmod L) + 1$ based on its depth $h$, and is represented as a pair $(\config, \ell)$. 
This index determines the level-specific action set used during the search. 
Unlike standard MCTS where the expansion stage considers the full set of possible exchanges, our approach restricts the expansion to the subset $\A^{(\ell)}(\config)\subseteq \A(\config)$, thereby enforcing a structured and focused exploration.
The complete algorithm is presented in the appendix, see Algorithm \ref{algorithm:ML-MCTS}.

Thanks to the mutually disjoint decomposition, the proposed ML-MCTS algorithm guarantees that all nodes at the same level and along the same path represent distinct configurations.
This resolves the redundancy issue in the standard MCTS and therefore improves search efficiency. 
In addition, the action space for a node at level $\ell$ is restricted to a much smaller action set $\A^{(\ell)}(\config)$, which progressively narrows the search scope from global to local as the level increases.

\begin{figure}[htb!]
\centering
\includegraphics[width=15.8cm]{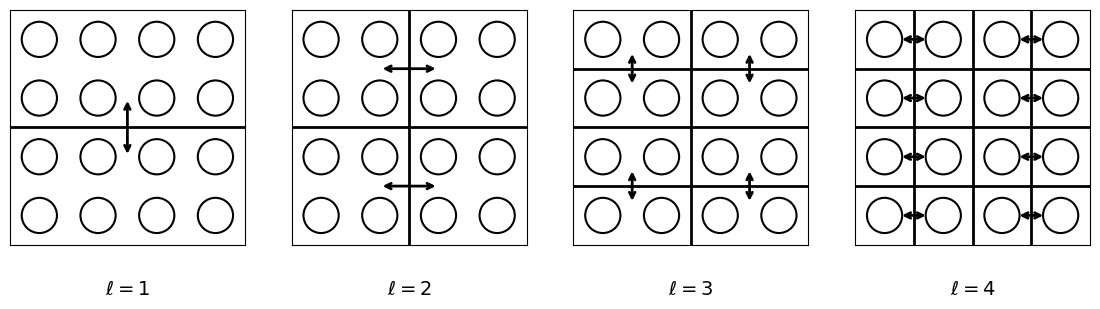}
\caption{Illustration of the multi-level construction for a two-dimensional square lattice with $4 \times 4$ sites in the supercell. Hollow circles represent lattice sites without specifying atomic species, and double-headed arrows indicate allowed exchange positions at each level.}
\label{fig-decompostion}
\end{figure}

The easiest and most natural hierarchical decomposition for crystalline system could be obtained with bisection.
We present an example in Figure \ref{fig-decompostion} with a two-dimensional $4 \times 4$ lattice, where the allowed permutations at each level are indicated by double-headed arrows.

\section{Numerical experiments}
\label{sec:numerics}
\setcounter{equation}{0}
\setcounter{figure}{0}

In this section, we present some numerical experiments on two-dimensional hexagonal lattice systems to demonstrate the efficiency of our ML-MCTS algorithm.
We consider the supercell configuration of a doped graphene system, with a number of atomic sites are substituted by other atomic species.
The objective function $f$ is modeled by the total energy of the system given by an empirical potential, whose formulations and parameters given in \ref{append:tersoff}.

Since the primary computational cost during the search process comes from evaluating the objective functions on different configurations, we assess the efficiency of each algorithm by tracking the decay of $f$-values with respect to the number of $f$-evaluations.
We compare the performance of three different methods: MH, MCTS, and ML-MCTS.
All simulations are performed on a mac with an Apple M1 Pro processor (10-core, 3.2 GHz) and 16 GB of RAM.
All numerical results are reported in atomic units (a.u.).

\vskip 0.3cm

\noindent
\textbf{Example 1 (Si-doping).}
We consider a graphene system doped with silicon atoms.
We take a supercell containing $8 \times 8$ unit cells, with $N=128$ lattice sites in $\Lambda$. 
The number of silicon atoms is chosen as 8, 12, and 16.

\begin{figure}[htbp!]
\centering
\subfigure[8 Si-doping]{
\includegraphics[width=0.31\textwidth]{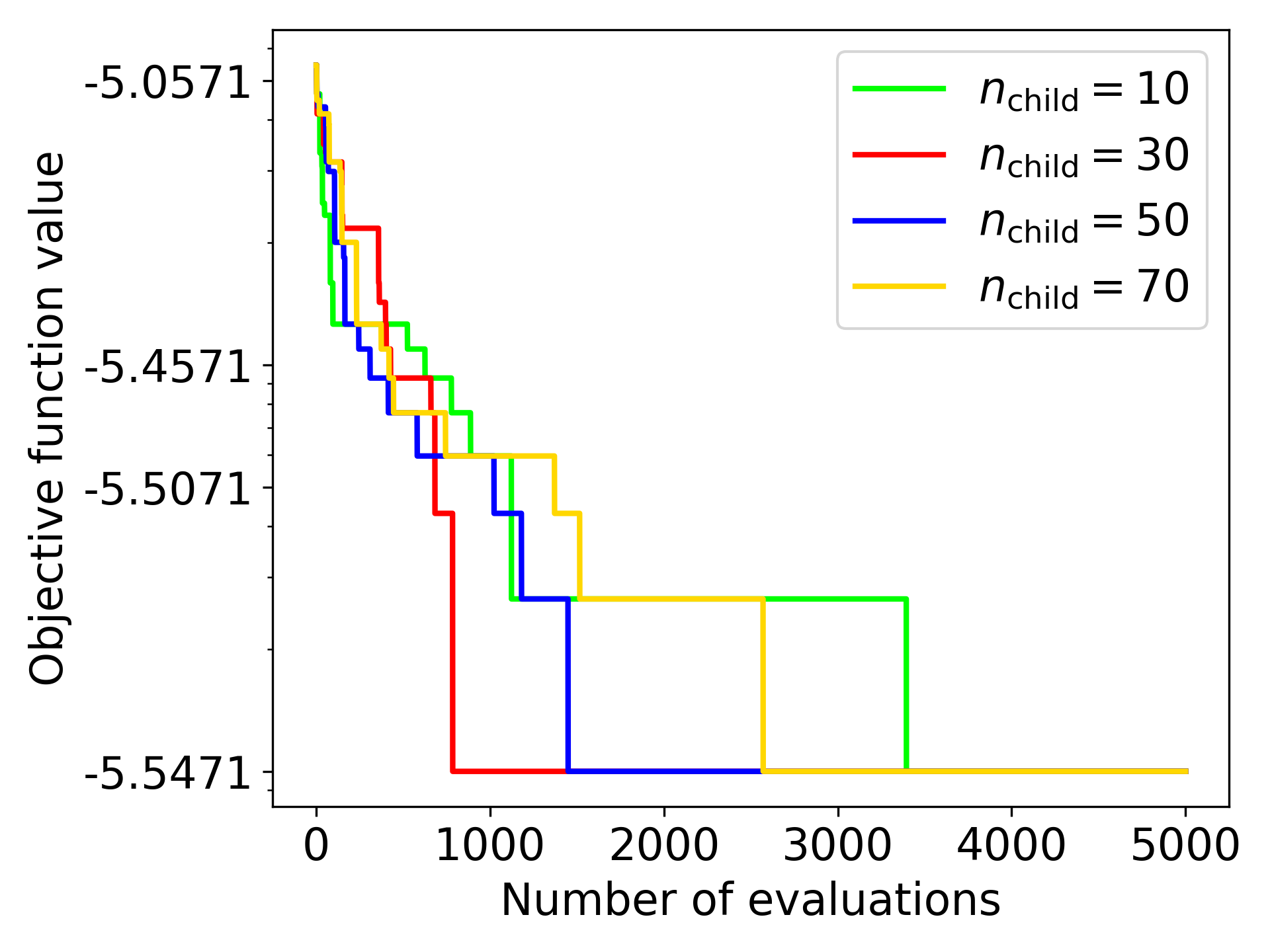}
}
\subfigure[12 Si-doping]{
\includegraphics[width=0.31\textwidth]{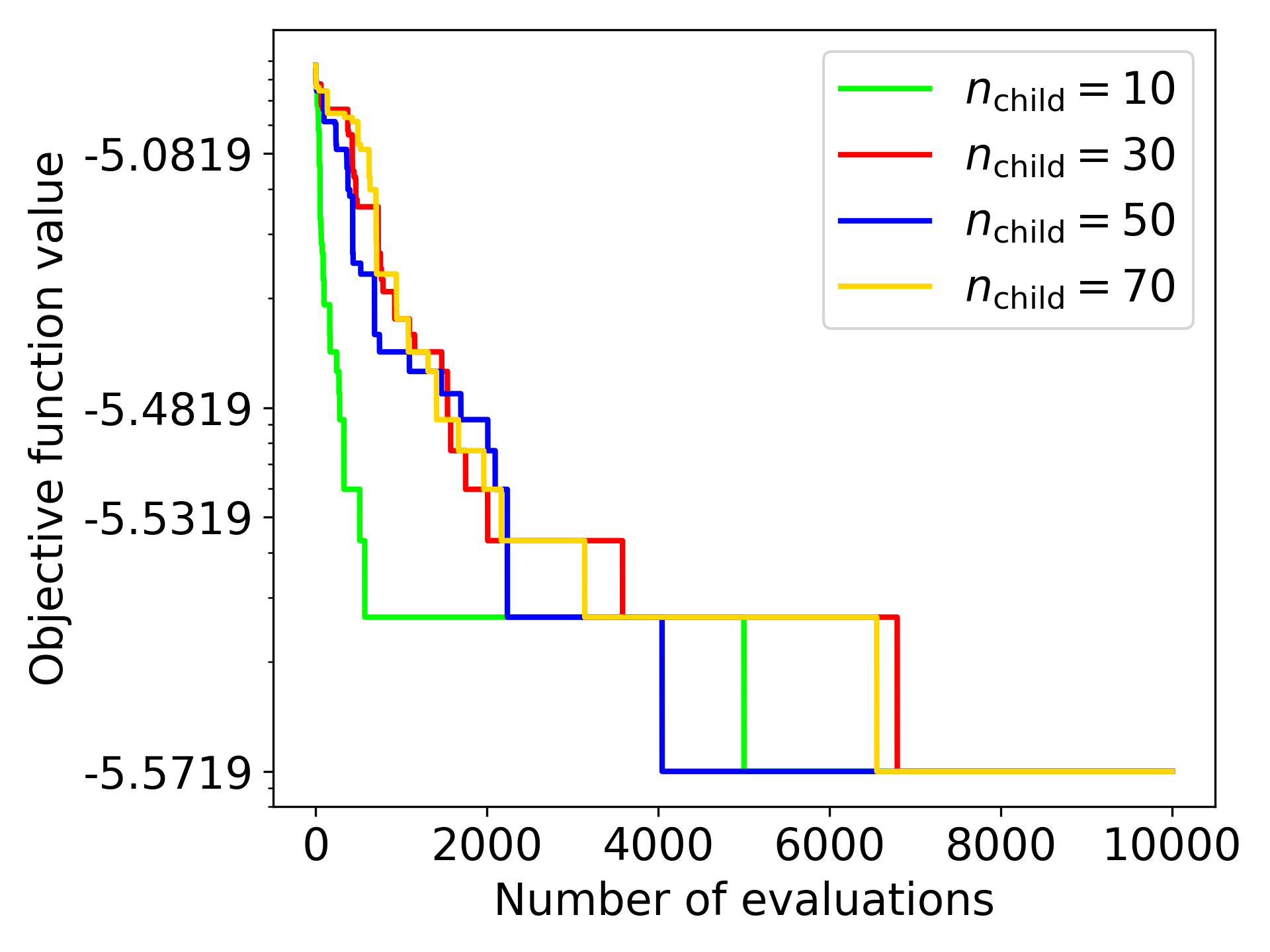}
}
\subfigure[16 Si-doping]{
\includegraphics[width=0.31\textwidth]{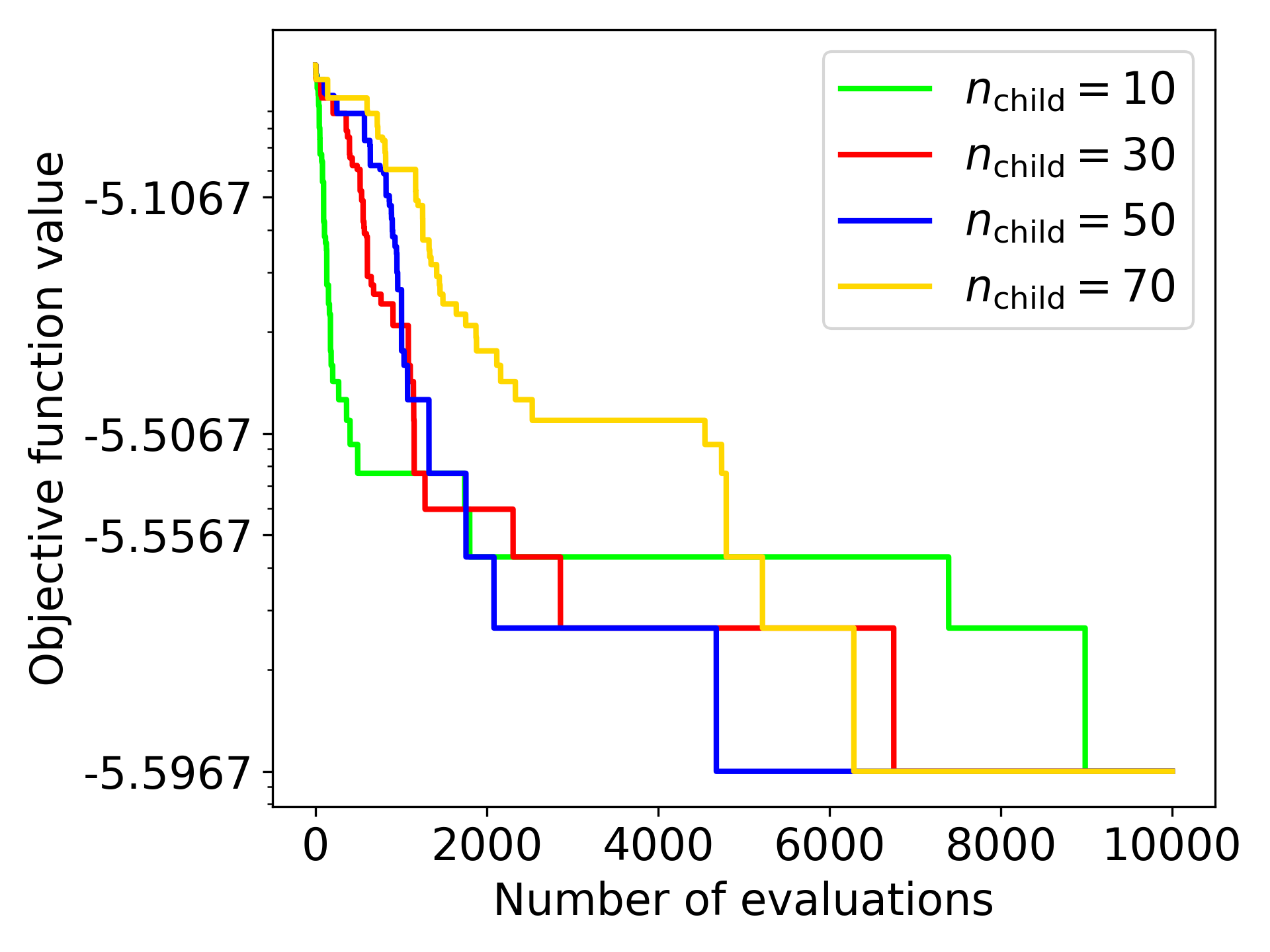}
}
\caption{Decay of the objective functions obtained with different $n_{\rm child}$.
}
\label{fig:nchild_converge}
\vskip 0.3cm
\centering
\subfigure[8 Si-doping]{
\includegraphics[width=0.31\textwidth]{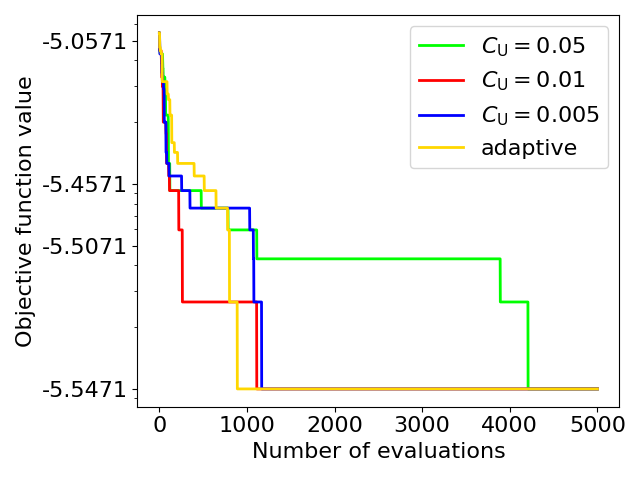}
}
\subfigure[12 Si-doping]{
\includegraphics[width=0.31\textwidth]{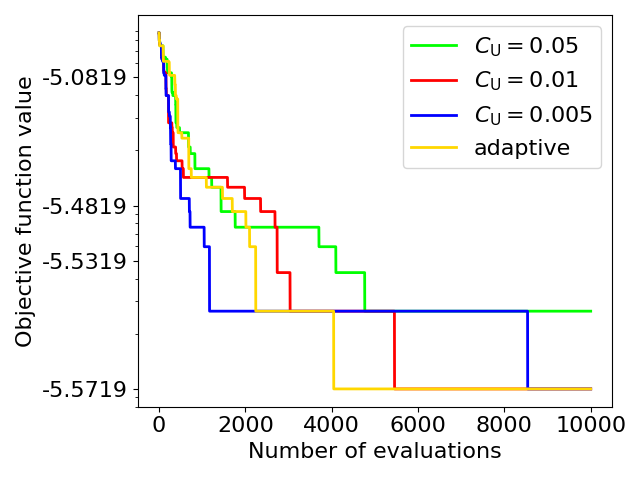}
}
\subfigure[16 Si-doping]{
\includegraphics[width=0.31\textwidth]{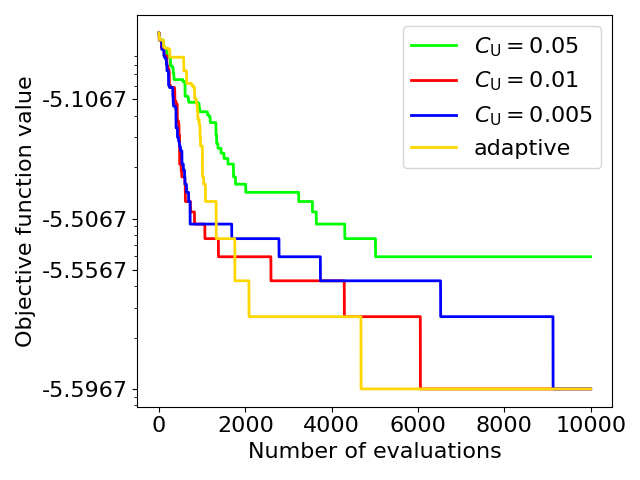}
}
\caption{Decay of the objective functions with different choices of $C_{\rm U}$.
}
\label{fig:ratio_converge}
\vskip 0.3cm
\centering 
  \subfigure[$8$ Si-doping]
{\includegraphics[width=0.32\textwidth]{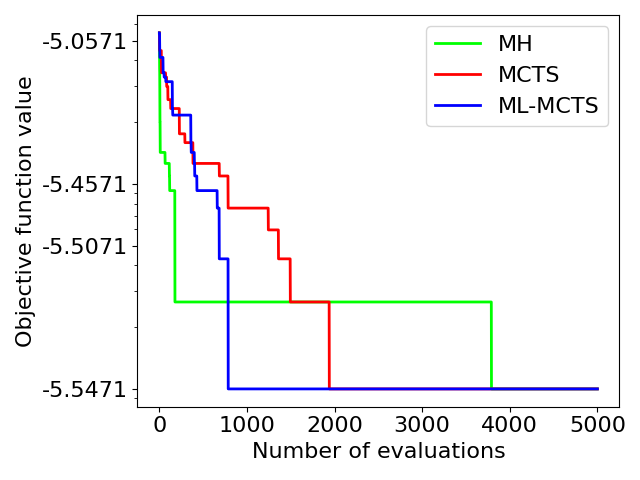}}
  \subfigure[$12$ Si-doping]
{\includegraphics[width=0.32\textwidth]{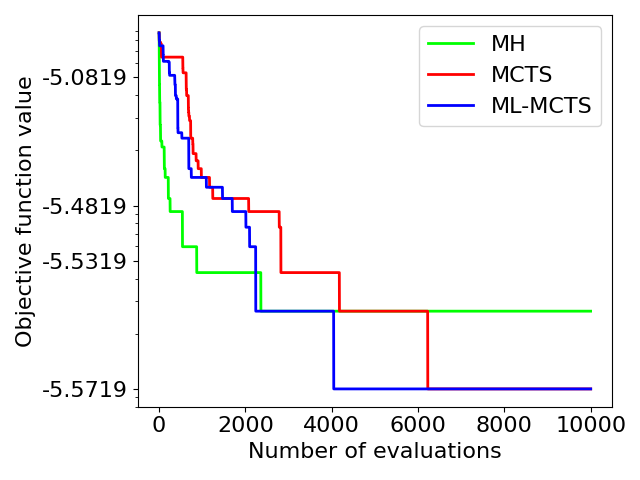}} 
  \subfigure[$16$ Si-doping]
{\includegraphics[width=0.32\textwidth]{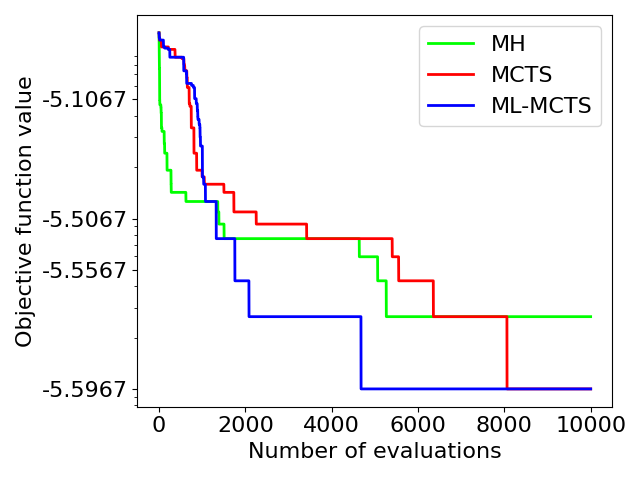}}
\caption{Decay of objective functions over $8\times 8$ unit cell configurations.
}
\label{fig:convergence}
\end{figure}

We first test the effects of two hyperparameters in the MCTS method: the maximum number of child nodes per node $n_{\rm child}$, and the exploration–exploitation trade-off parameter $C_{\rm U}$ in \eqref{def:exploration}.
In Figure \ref{fig:nchild_converge}, we show the decay of $f$-values under different choices of $n_{\rm child}$, from which we can observe that an optimal choice of $n_{\rm child}$ increases with respect to the number of doping atoms.
The reason could be that both the configuration space and the corresponding action space increases significantly with respect to the number of doping atoms.
In Figure \ref{fig:ratio_converge}, we compare different strategies for selecting the exploration–exploitation parameter $C_{\rm U}$, with fixed values and an adaptive adjustment.
In the adaptive strategy, the parameter $C_{\rm U}$ is set with respect to the number of past $f$-evaluations $k$
\begin{equation*}
C_{\rm U} = \max \big\{ 0.1 \times \exp(-0.001 \times k), 0.01 \big\} ,
\end{equation*}
which gradually shifts the focus from exploration to exploitation as the search progresses.
We can observe from the picture that the adaptive strategy outperforms the others with fixed $C_{\rm U}$.

\begin{figure}[htbp!]
\vskip -0.5cm
\centering
\includegraphics[width=1.0\textwidth]{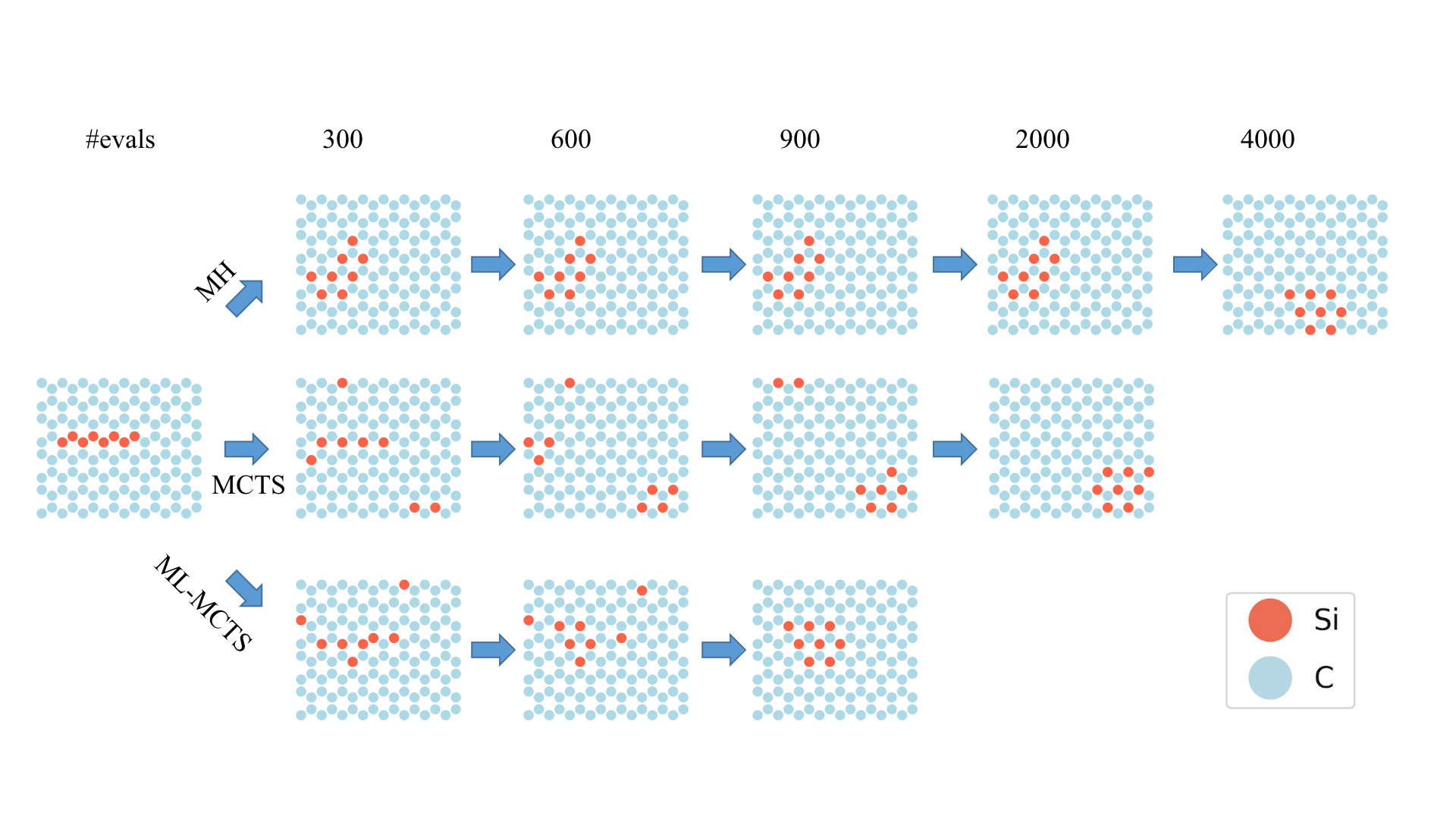}
\vskip -0.5cm
\caption{Evolution of configurations with $8$ Si-doping during the search process obtained by different methods.
The numbers on the first row indicate the number of evaluations at which the configurations were obtained.
The evolution corresponds to the objective function decay in Figure \ref{fig:convergence}(a).
}
\label{fig:search-process-8}
%
\vskip -0.1cm
%
\centering
\includegraphics[width=1.0\textwidth]{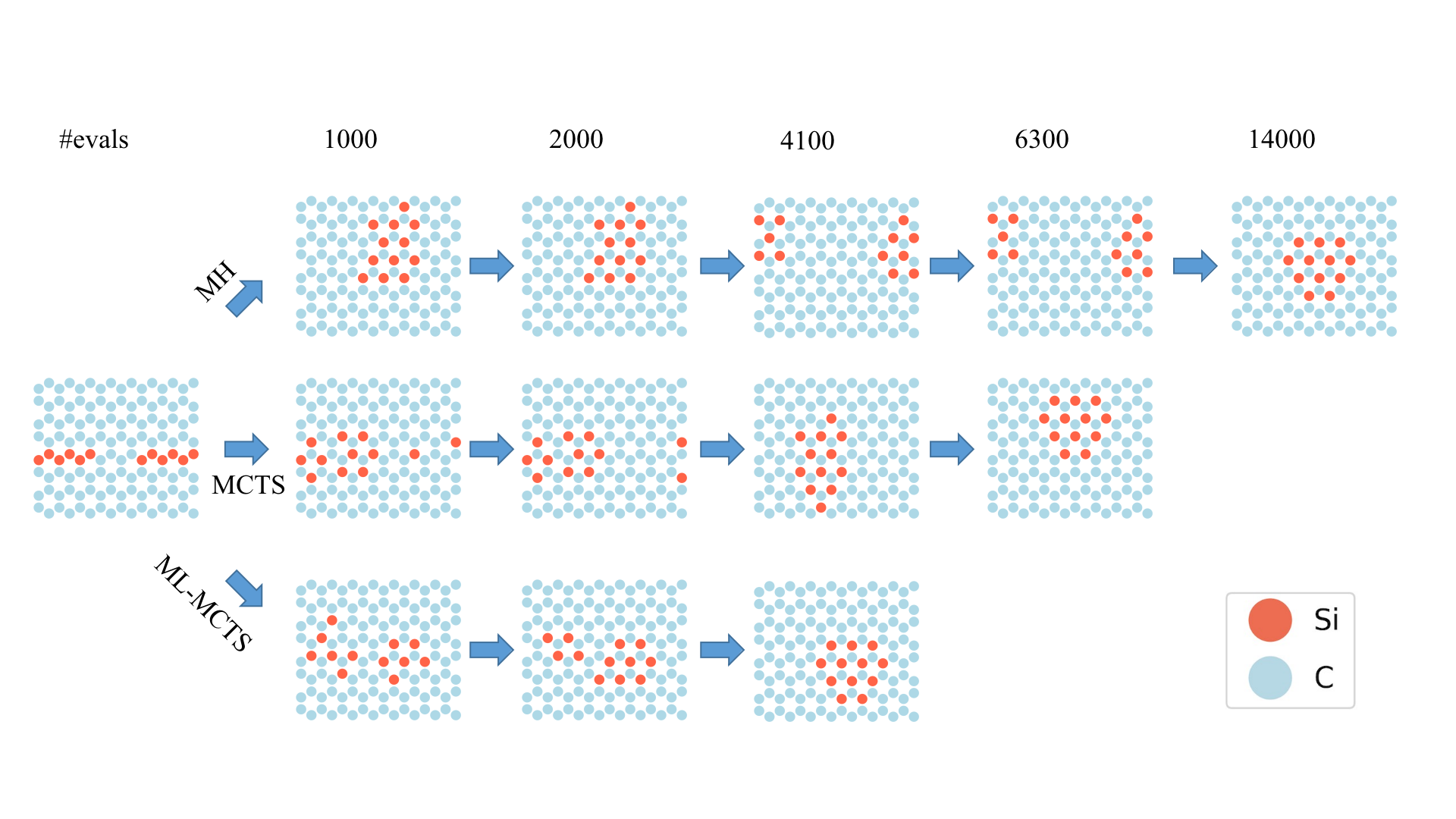}
\vskip -0.5cm
\caption{Evolution of configurations with $12$ Si-doping during the search process obtained by different methods.
The numbers on the first row indicate the number of evaluations at which the configurations were obtained.
The evolution corresponds to the objective function decay in Figure \ref{fig:convergence}(b).
}
\label{fig:search-process-12}
\end{figure}

We then compare the performance of three different search methods across various doping ratios. 
We observe that the ML-MCTS method exhibits the fastest decay among them, followed by standard MCTS and then the MH method, which aligns with our expectations.
The results also show that the MH method tends to achieve rapid descent during the initial phase of the search, but often stagnates or becomes trapped near certain values (corresponding to some local minima).
In contrast, the MCTS and ML-MCTS methods are able to escape from local minima due to its guided selection mechanism. 
Moreover, as the number of doping silicon atoms increases, the configuration space becomes larger very quickly, and the advantage of the ML-MCTS becomes more pronounced, demonstrating its scalability and robustness for complex search problems.
We finally compare the optimal configurations found at different stages of the search process, and present the systems with 8 silicon atoms in Figure \ref{fig:search-process-8} and systems with 12 silicon atoms in Figure \ref{fig:search-process-12}.
We observe that all three methods can find the same (equivalent) optimal configurations at the end, while the MH and MCTS methods get the finial solution with many more $f$-evaluations than the ML-MCTS method, as they tend to be trapped at some local minima during the search process.

\vskip 0.3cm

\begin{figure}[htb!]
  \centering 
  \subfigure[$\rm{B}_3 \rm{N}_3$-doping]
{\includegraphics[width=0.3\textwidth]{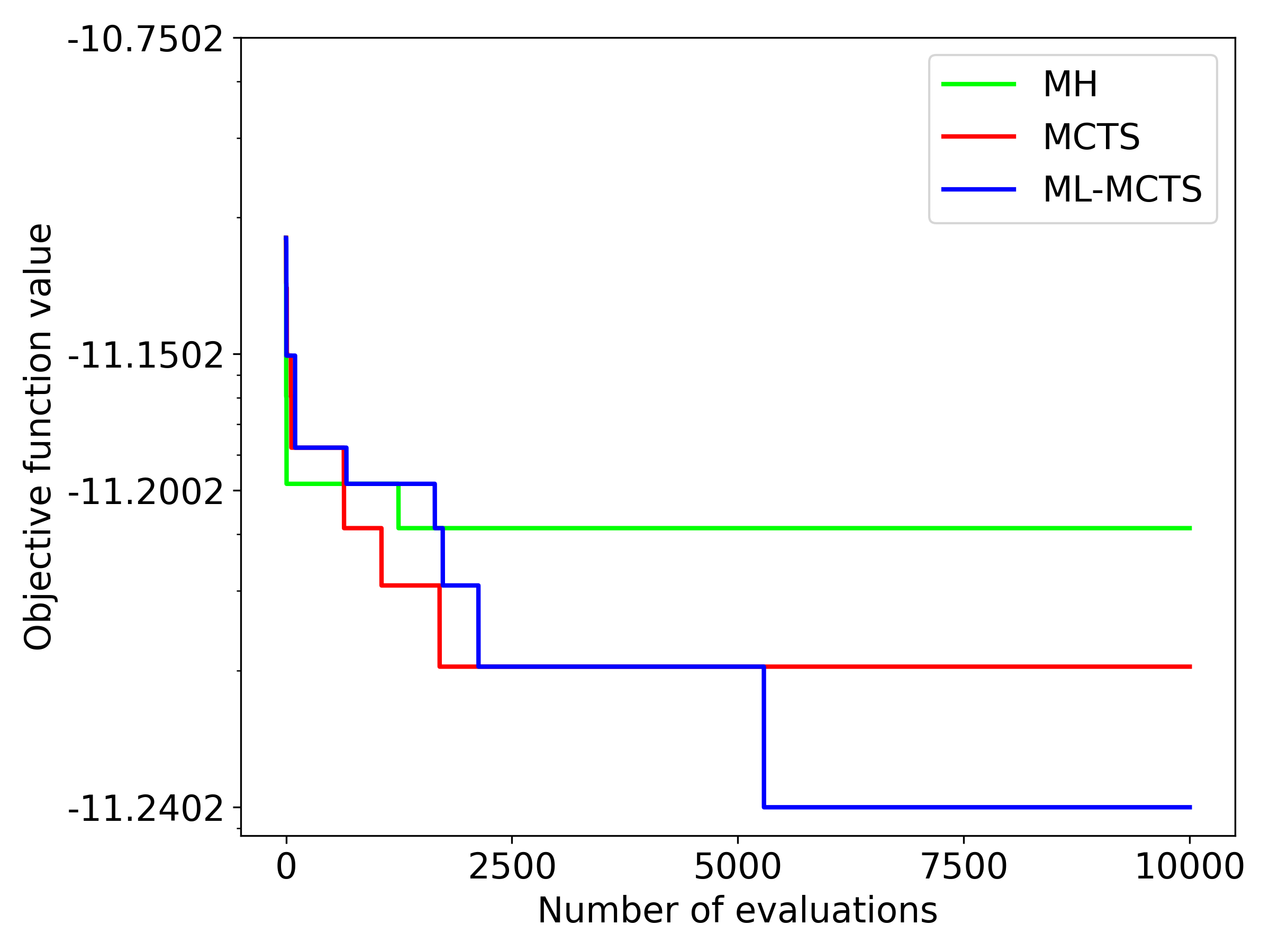}}
  \subfigure[$\rm{B}_3 \rm{N}_6$-doping]
{\includegraphics[width=0.3\textwidth]{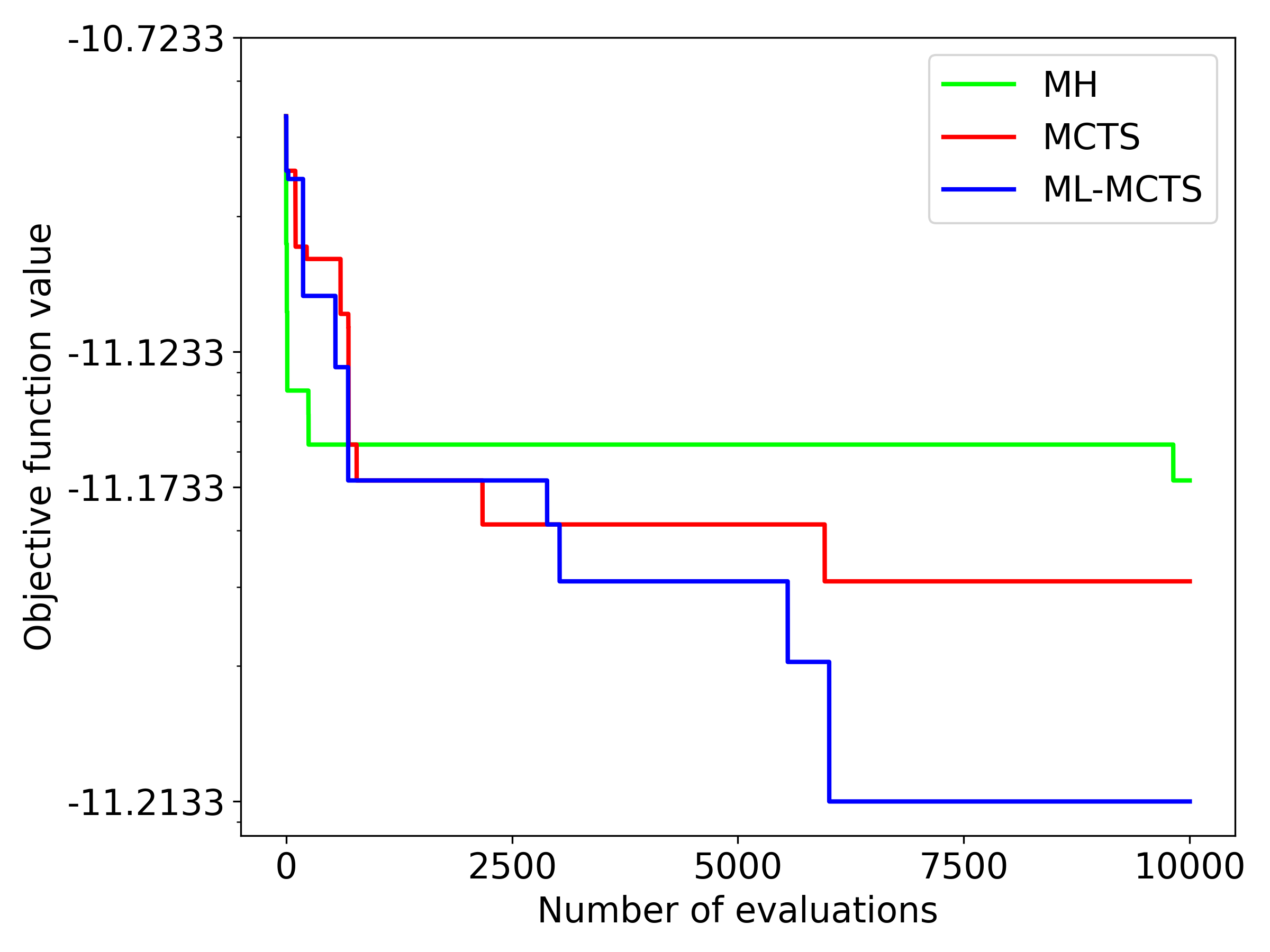}} 
  \subfigure[$\rm{B}_8 \rm{N}_8$-doping]
{\includegraphics[width=0.3\textwidth]{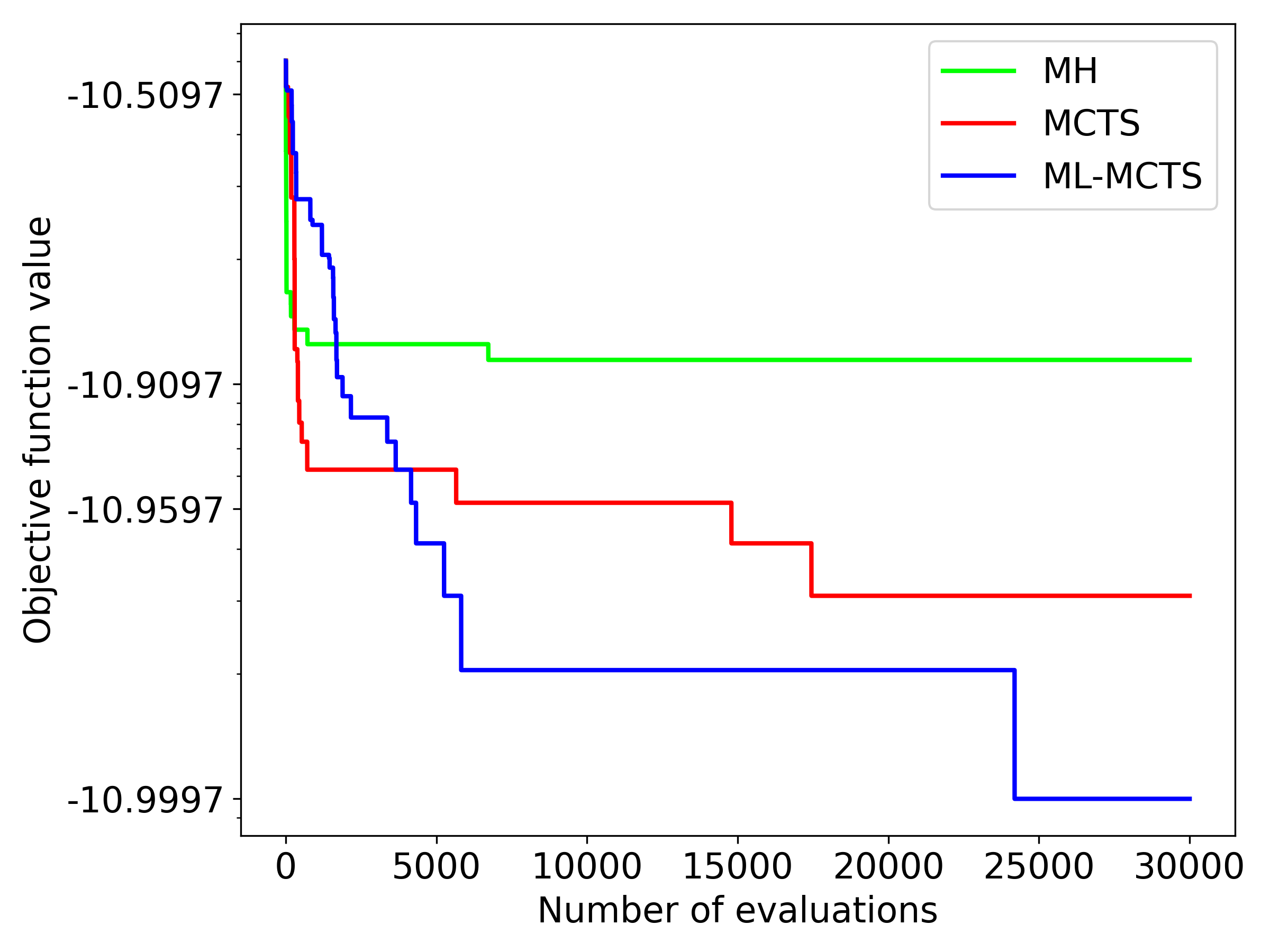}}
\caption{Decay of objective functions over $8\times 8$ unit cell configurations.
}
\label{fig:convergenceBN}
\end{figure}

\noindent
\textbf{Example 2 (BN-doping).} 
We investigate the doping of graphene systmes with boron and nitrogen atoms.
Boron and nitrogen are adjacent to carbon in the periodic table, and when introduced in pairs, they preserve the overall valence electron count of the system.
Such charge-neutral co-doping schemes are of particular interest for designing semiconducting or insulating derivatives of graphene \cite{ci2010atomic, rubio2010nanoscale}. 
In our experiments, we consider both balanced co-doping cases (equal numbers of B and N atoms) and unbalanced cases where the numbers differ.
In Figure \ref{fig:convergenceBN}, we compare the performance of three different search methods under these settings.
In all cases, the ML-MCTS method consistently shows the fastest decay in the objective value, demonstrating superior search efficiency.

\begin{figure}[htb!]
\centering 
\subfigure[MH]
{\includegraphics[width=0.3\textwidth]{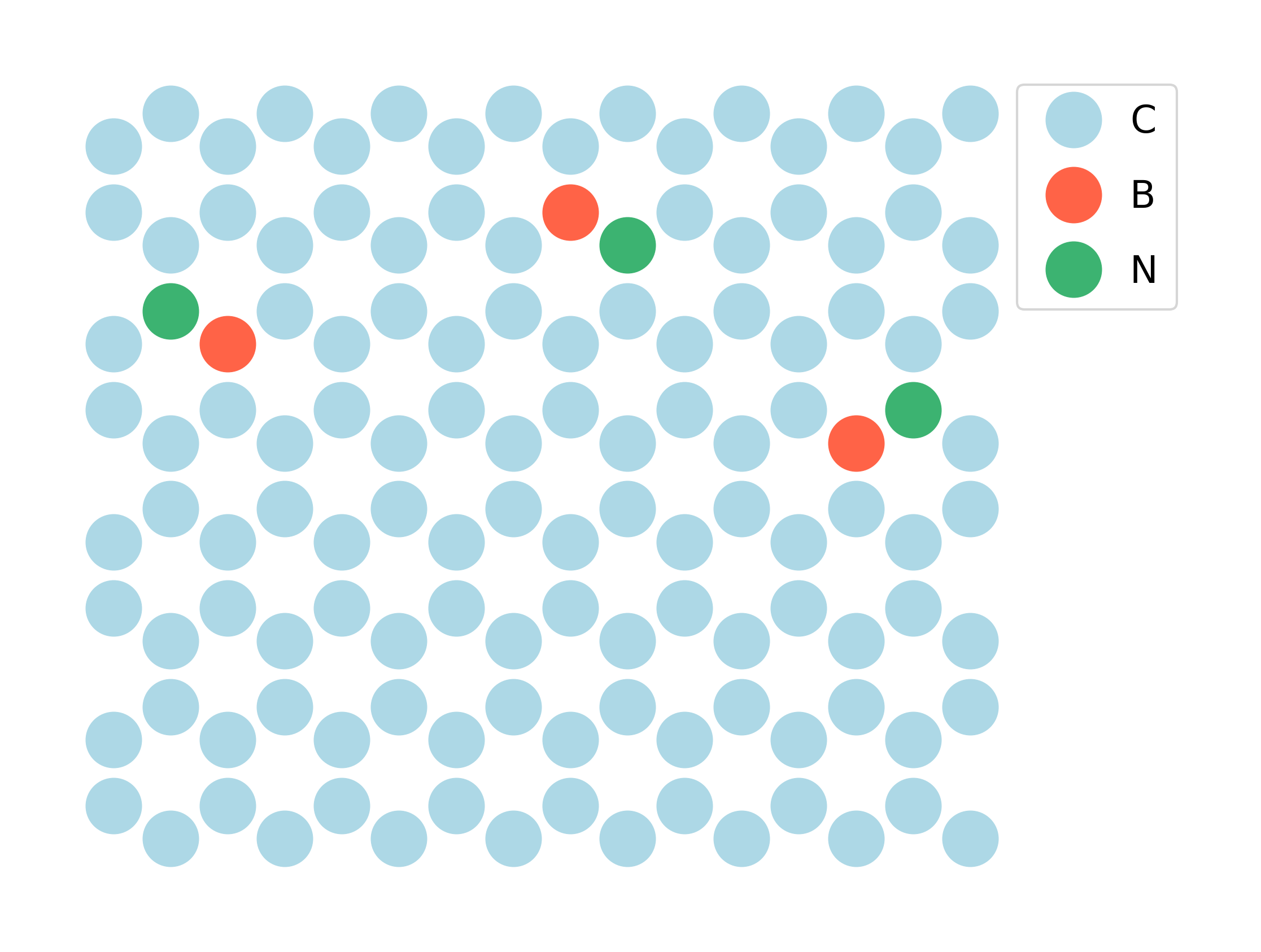}}
  \subfigure[MCTS]
{\includegraphics[width=0.3\textwidth]{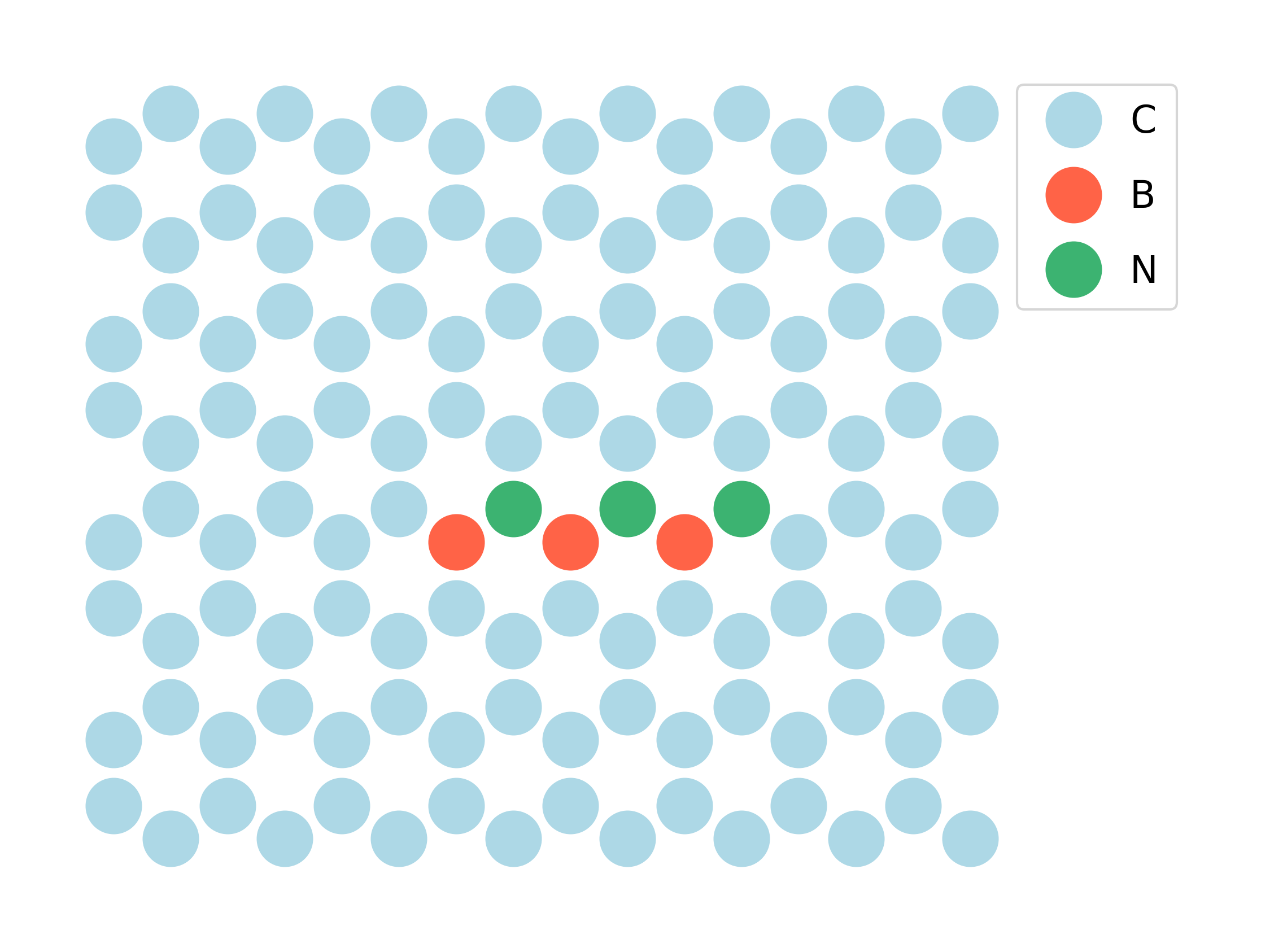}} 
  \subfigure[ML-MCTS]
{\includegraphics[width=0.3\textwidth]{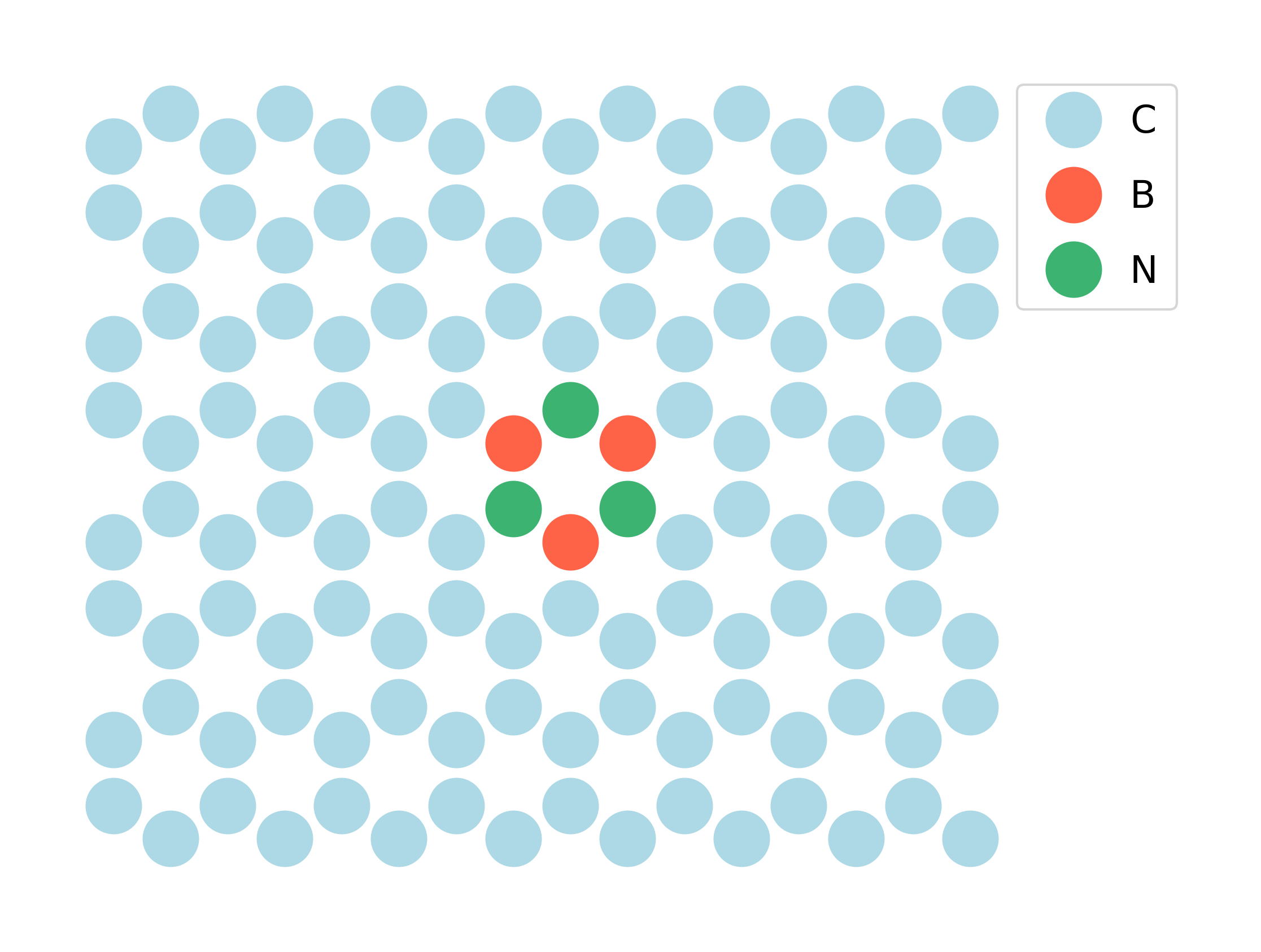}}
\caption{Configurations obtained by three different methods after $10000$ search iterations for $\rm{B}_3 \rm{N}_3$-doped graphene.
}
\label{fig:configBN-dop3}
\vskip 0.3cm
\centering 
\subfigure[MH]
{\includegraphics[width=0.3\textwidth]{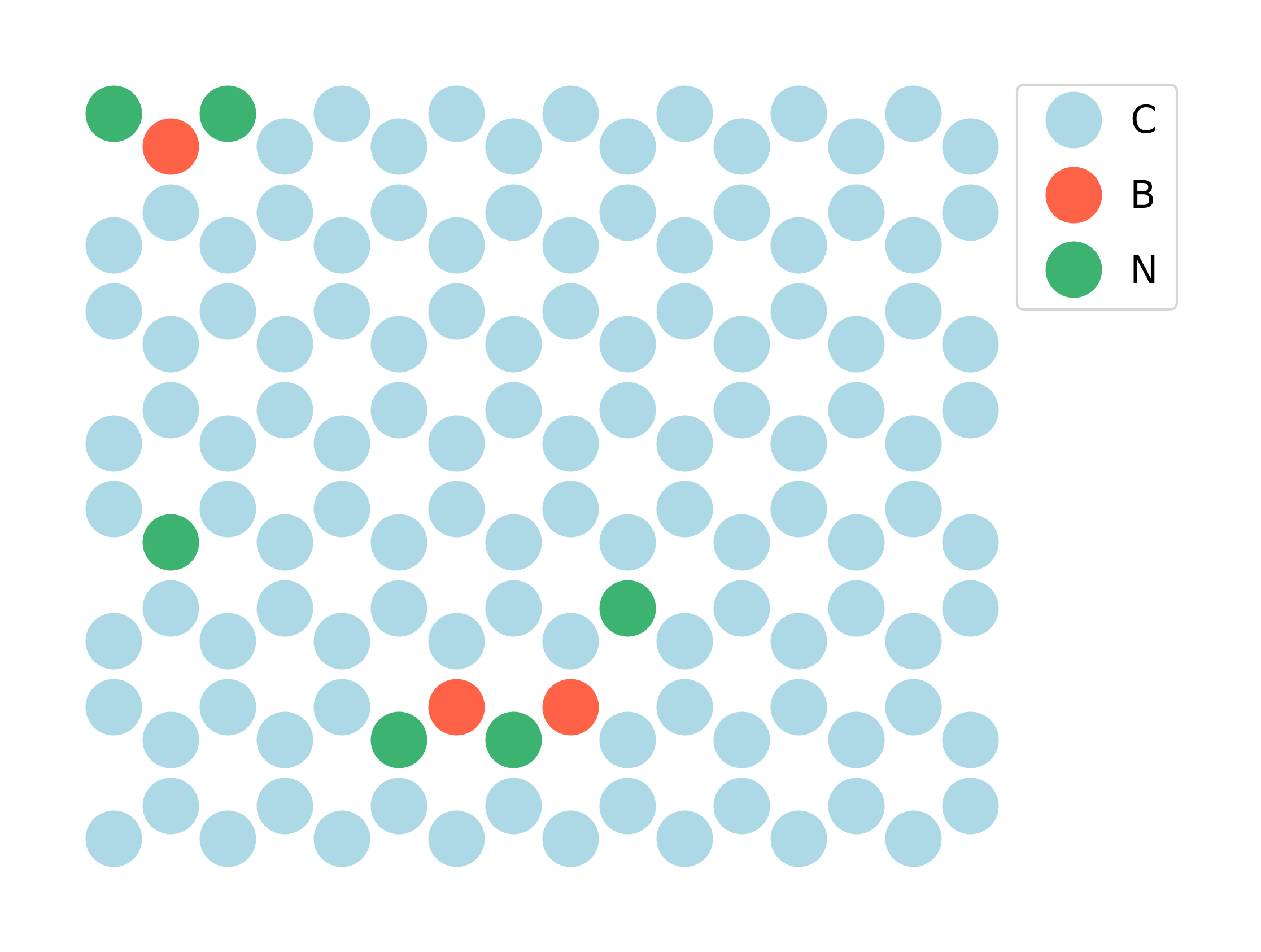}}
  \subfigure[MCTS]
{\includegraphics[width=0.3\textwidth]{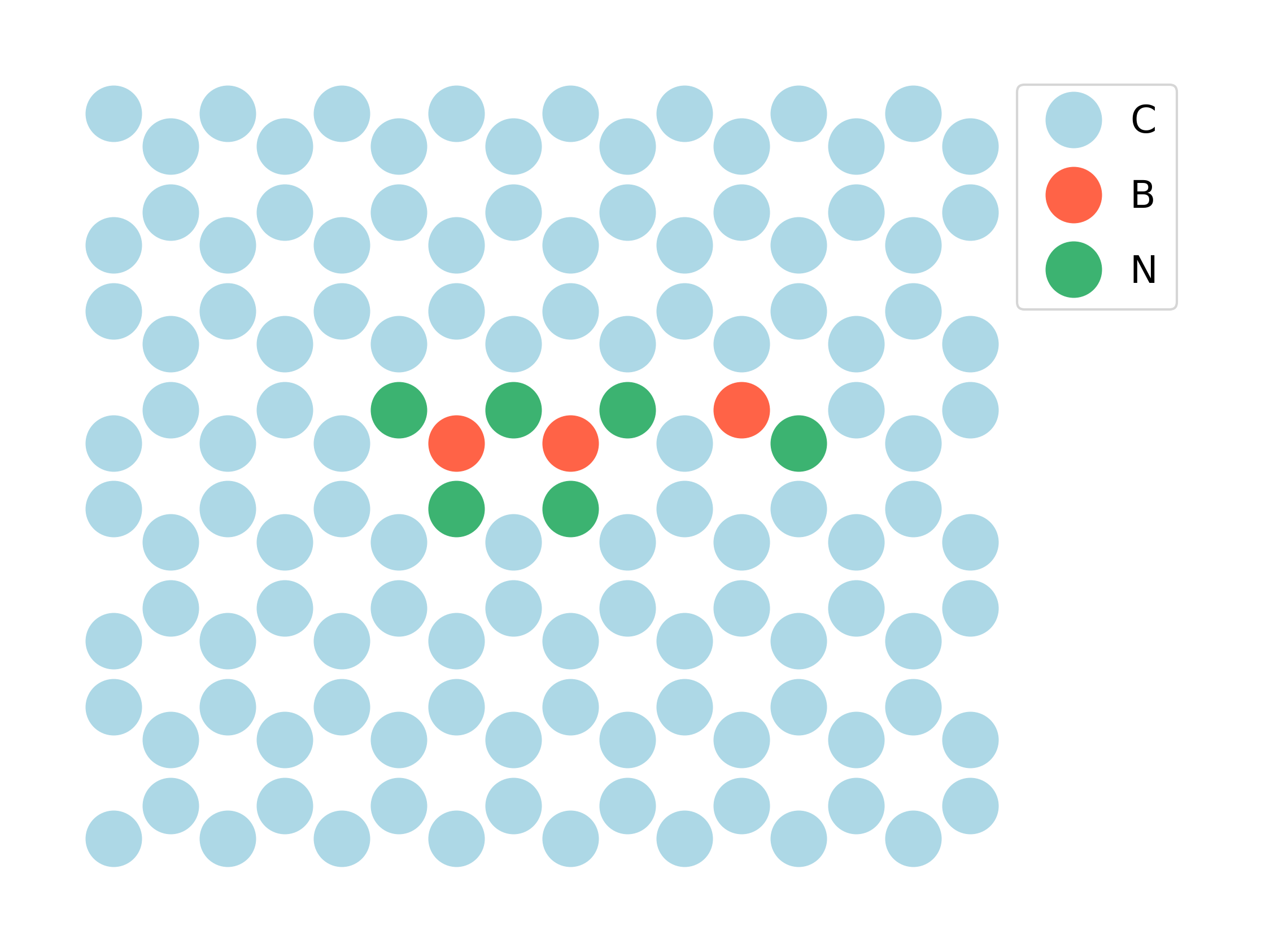}} 
  \subfigure[ML-MCTS]
{\includegraphics[width=0.3\textwidth]{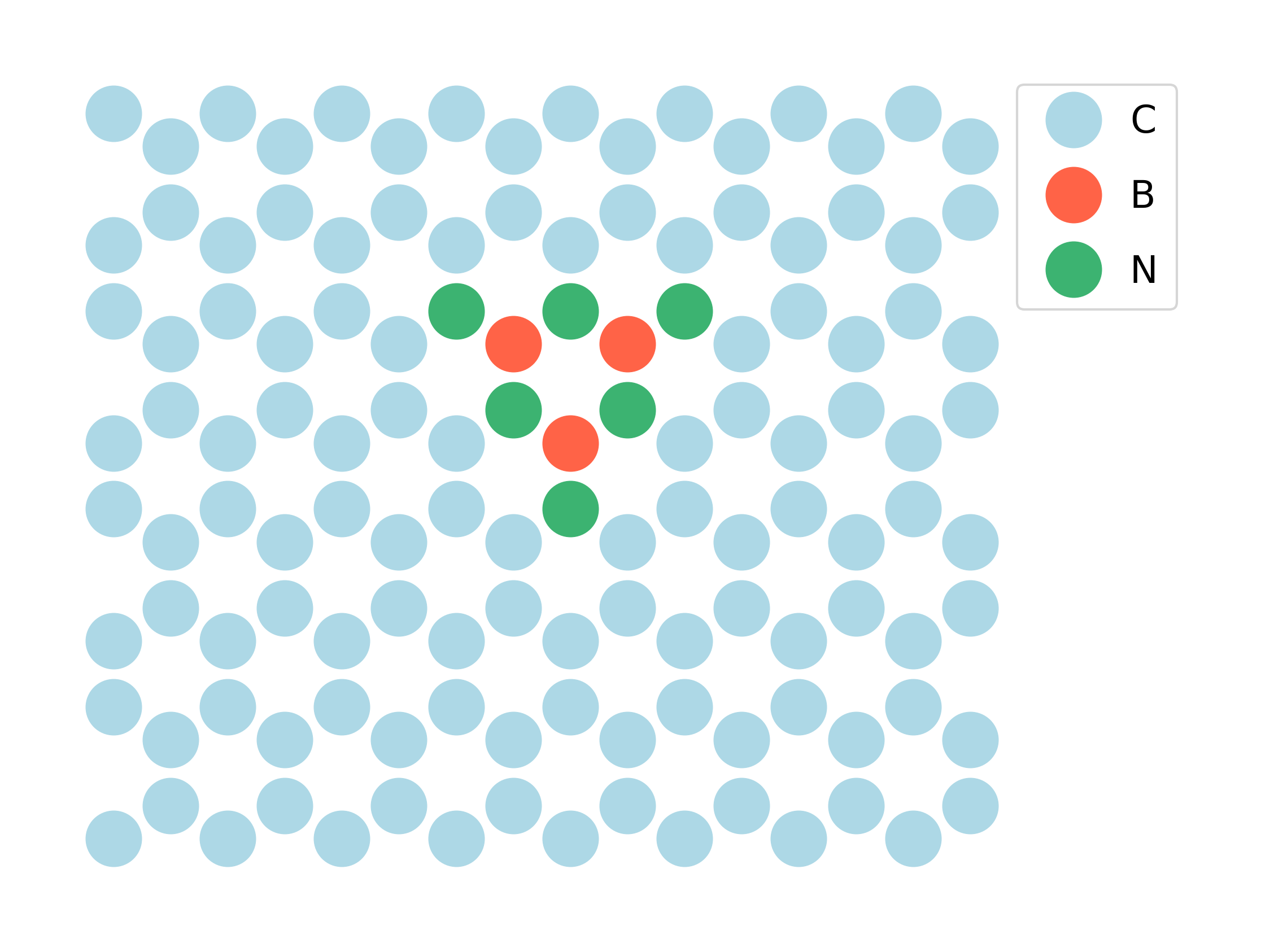}}
\caption{Configurations obtained by three different methods after $10000$ search iterations for $\rm{B}_3 \rm{N}_6$-doped graphene.
}
\label{fig:configBN-dop5}
\vskip 0.3cm
\centering 
\subfigure[MH]
{\includegraphics[width=0.3\textwidth]{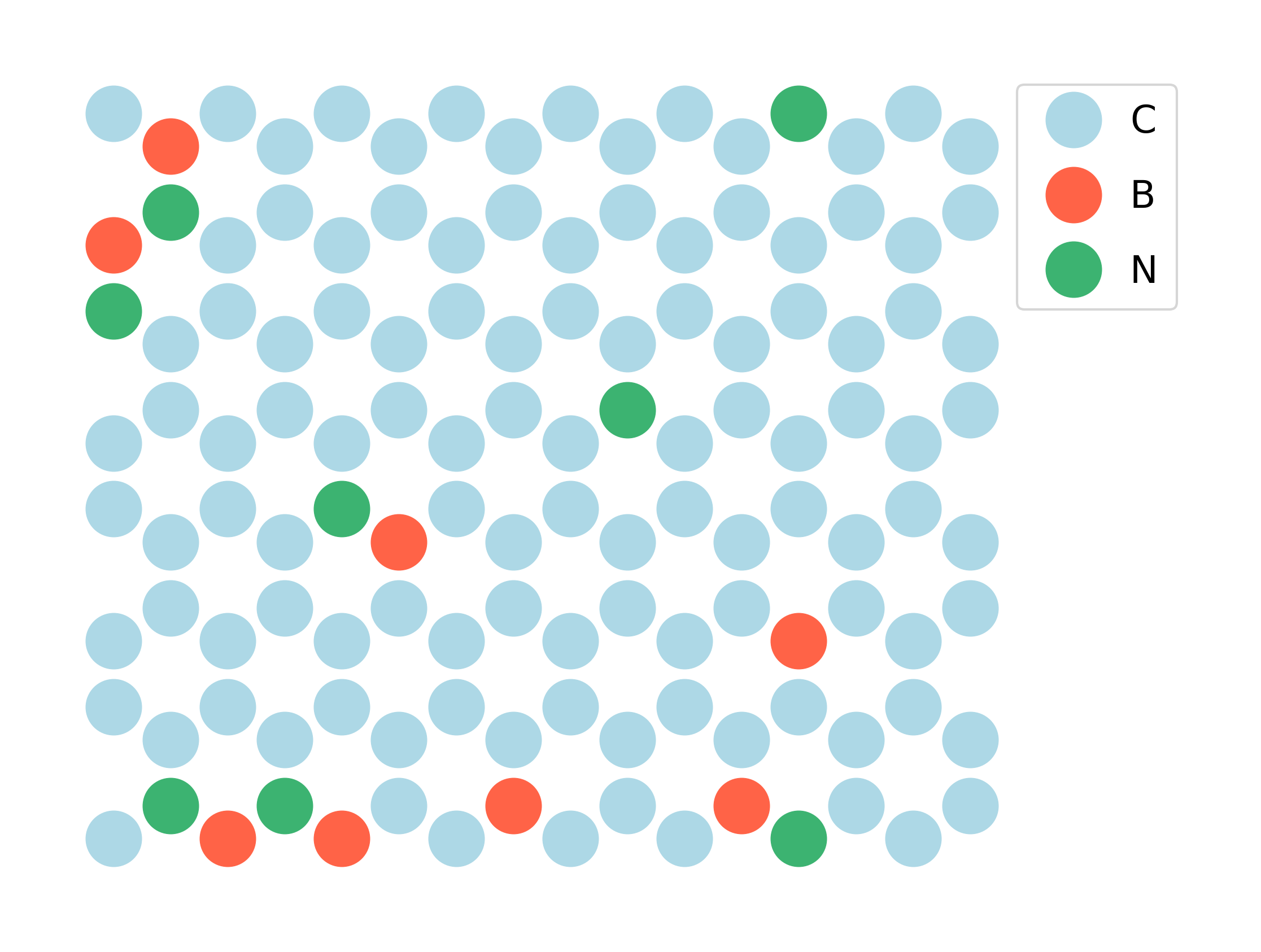}}
\subfigure[MCTS]
{\includegraphics[width=0.3\textwidth]{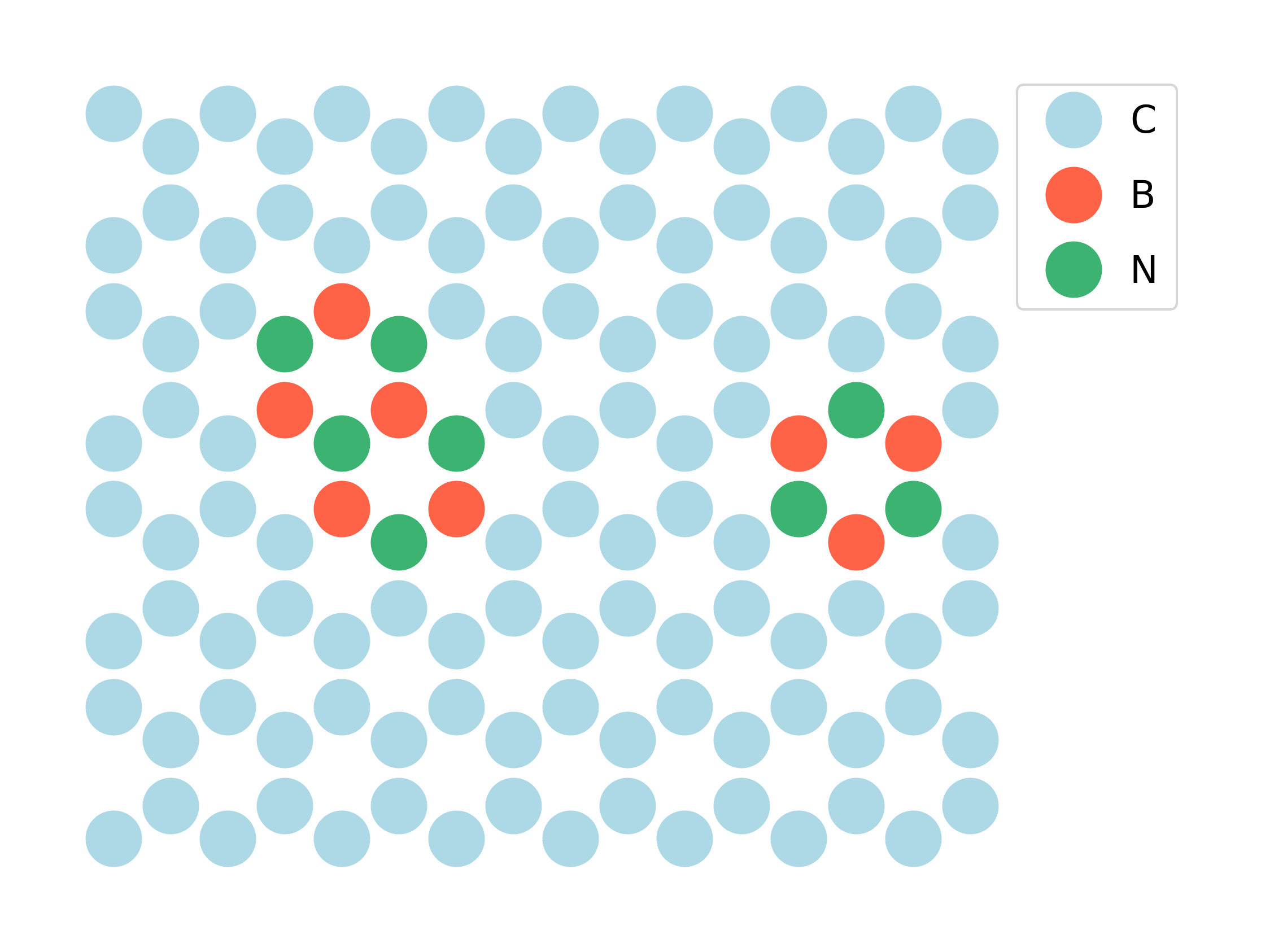}} 
\subfigure[ML-MCTS]
{\includegraphics[width=0.3\textwidth]{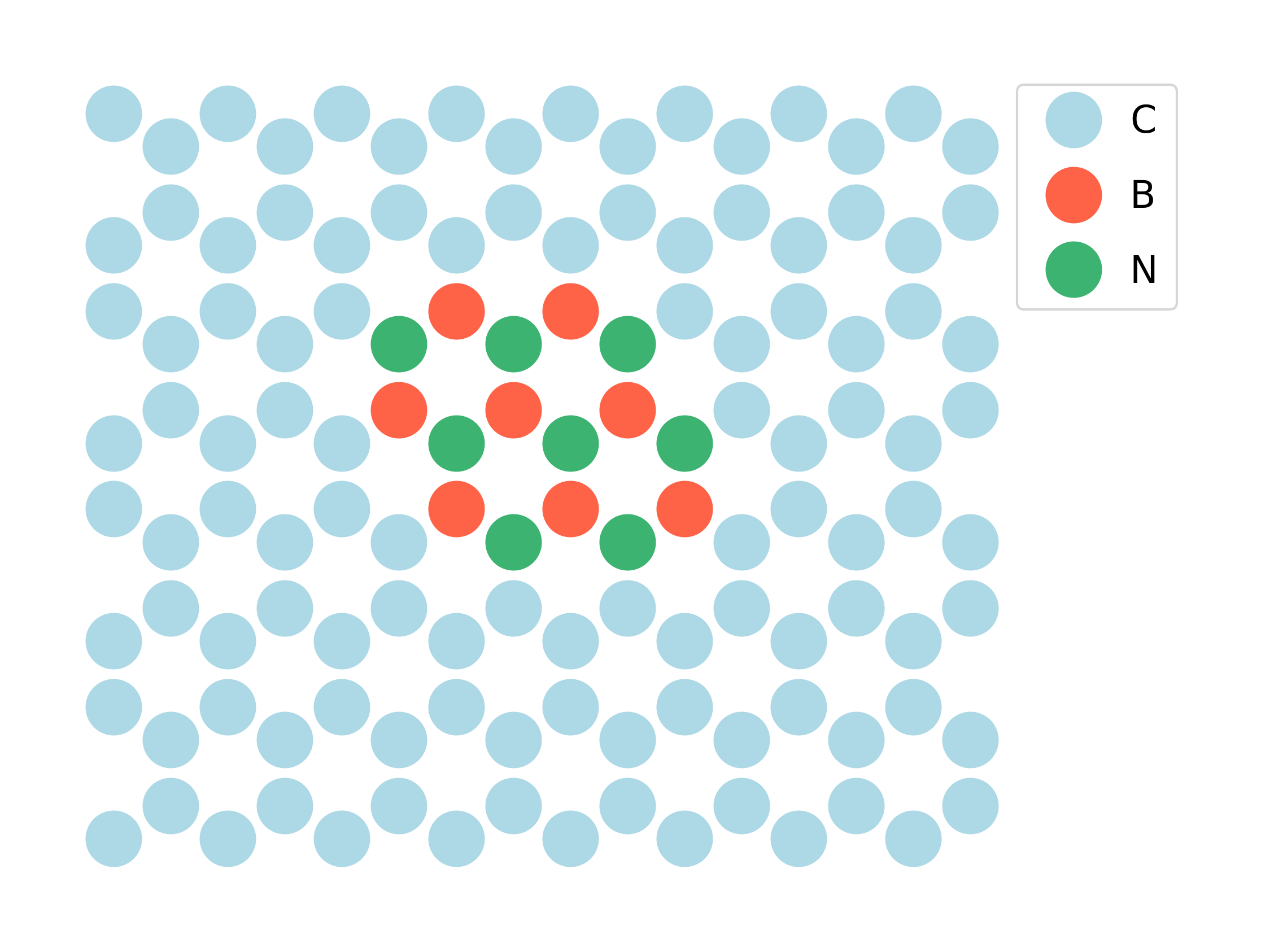}}
\caption{Configurations obtained by three different methods after $30000$ search iterations for $\rm{B}_8 \rm{N}_8$-doped graphene.
}
\label{fig:configBN-dop8}
\end{figure}

In Figure \ref{fig:configBN-dop3}, Figure \ref{fig:configBN-dop5} and Figure \ref{fig:configBN-dop8}, we show the optimal configurations within a given number of $f$-evaluations for three different methods.
We observe that although no explicit constraint is imposed during the doping process to enforce the formation of B–N bonded pairs, the optimal configurations identified by our ML-MCTS algorithm naturally exhibit such pairing behavior.
This result aligns with typical strategies in materials modeling \cite{fan2012band, shinde2011direct}, where boron nitride (BN) domains are often embedded into graphene to tune its electronic properties and improve conductivity control.
This result highlights the strength of our method in efficiently discovering physically meaningful configurations without relying on manually imposed constraints or heuristic assumptions.
In contrast, the MH and standard MCTS methods often converge to suboptimal configurations, 
especially under the same number of $f$-evaluations, frequently missing the key features such as B–N pairing that are expected from physical and chemical considerations.
Such configurations may correspond to local minima that are less relevant for realistic material behavior.

Efficient identification of low-energy and stable atomic structures is essential in materials science, as these configurations determine not only thermodynamic feasibility but also critical material properties such as bandgap, conductivity, and chemical reactivity.
In the case of B/N-doped graphene, the stability of dopant arrangements directly affects its potential applications in nanoelectronics, sensors, and catalytic systems \cite{dong2024electron, tran2015high, wang2018thermal}.
Our results indicate that ML-MCTS can serve as a useful tool for exploring stable atomic configurations in doped systems, thereby assisting in the understanding and design of materials with targeted properties.

\section{Conclusions}
\label{sec:conclusion}

In this work, we propose a novel ML-MCTS method for the combinatorial optimizaiton problem arising in configuration generation for crystalline systems. 
By decomposing the action space across hierarchical levels and progressively refining the configuration landscape, our method significantly reduces the tree search redundancy and enables the discovery of optimal configurations with high efficiency.
Numerical experiments on some doped graphene systems show that our ML-MCTS method consistently outperforms traditional MH and standard MCTS approaches in both convergence speed and final solution quality.
In future work, we plan to extend this method to the construction of disordered configurations in more complex alloy systems. These systems often involve multiple atomic species and additional physical constraints, where efficient and scalable search algorithms like ML-MCTS can play a key role in identifying low-energy structures and guiding materials design.

\section*{Acknowledgement}

This work was supported by the National Key R\&D Program of China (No. 2020YFA0712900).
The first author was partially supported by the National Natural Science Foundation of China (No. 12301548) and the Open Project Program of Key Laboratory of Mathematics and Complex System (Grant No. K202302), Beijing Normal University.
The third author was partially supported by the National Natural Science Foundation of China (No. 12371431).
The fourth author was partially supported by National Natural Science Foundation of China under grants U23A20537 and Funding of National Key Laboratory of Computational Physics.

\appendix
\renewcommand\thesection{\appendixname~\Alph{section}}

\section{The ML-MCTS algorithm}
\label{append:algorithm}
\renewcommand{\theequation}{A.\arabic{equation}}
\renewcommand{\thetable}{A.\arabic{table}}
\renewcommand{\thealgorithm}{A.\arabic{algorithm}}
\setcounter{equation}{0}
\setcounter{table}{0}

In this appendix, we present the detailed ML-MCTS algorithm in Algorithm \ref{alg:ml-mcts}.

\begin{algorithm}[htb!]
\caption{Multi-Level Monte Carlo Tree Search (ML-MCTS)}
\label{alg:ml-mcts}
\begin{algorithmic}[1]
\State \textbf{Input:} Initial configuration $\config_0$, maximum level $L$, maximum iteration times $T$
\State \textbf{Output:} Best configuration
\vspace{0.5em}
\State Initialize root node with $\config_0$ at level $\ell = 1$
\While{number of iterations $< T$}
    \vspace{0.5em}
    \State \textbf{Selection:}
    \State Set current node $\config \gets \config_0$
    \While{$\config$ is expandable}
        \State Select a child $\config' \in \text{Children}(\config)$ by maximizing
        $Q(\config') + R(\config,\config')$
        \State Set $\config \gets \config'$ and $\ell\gets\ell+1$
    \EndWhile
    \vspace{0.5em}
    \State \textbf{Expansion}:
    \State Sample an allowed action $(p,q) \in \A^{(\ell)}(\config)$
    \State Generate a new configuration $\config' \gets \text{Swap}(\config, p, q)$
    \State Add $\config'$ as a child of $\config$ at level $\ell + 1$
    \vspace{0.5em}
    \State \textbf{Simulation}:
    \State Evaluate $f(\config')$ and set $\Delta \gets f(\config')$
    \vspace{0.5em}
    \State \textbf{Backpropagation}:
    \ForAll{nodes $\config$ on the selected path}
        \State $n(\config) \gets n(\config) + 1$
        \State $Q(\config) \gets \max\{Q(\config), -\Delta\}$
    \EndFor
\EndWhile
\end{algorithmic}
\label{algorithm:ML-MCTS}
\end{algorithm}

\section{The Tersoff potentials}
\label{append:tersoff}
\renewcommand{\theequation}{B.\arabic{equation}}
\renewcommand{\thetable}{B.\arabic{table}}
\setcounter{equation}{0}
\setcounter{table}{0}

In this appendix, we present the detailed inter-atomic potentials that are used for the objective functions $f$ in the numerical experiments.

The total energy of the system is a function of configuration $\config\in\Sset$, given by
\begin{equation}
\label{tersoff-SiG}
E(\config) = \frac{1}{2} \sum_{i\in\Lambda} \sum_{(i\neq) j\in\Lambda} f_{\rm C}(r_{ij};\sigma_i,\sigma_j) \Big(f_{\rm R}(r_{ij};\sigma_i,\sigma_j) + f_{\rm A}(r_{ij};\sigma_i,\sigma_j) \Big),  
\end{equation} 
where $r_{ij}$ denotes the periodic distance between lattice sites $i$ and $j$ in the supercell, and $\sigma_i$ corresponds to the occupancy of different atoms.
The energy function is a simplified form of the Tersoff potential \cite{tersoff88empirical, tersoff88new}, an empirical potential widely used to model the energy of simulated systems.

The cutoff function $f_{\rm C}$ is defined by
\begin{equation*}
f_{\rm C}(r;I,J) = 
\left\{
\begin{array}{ll}
1   &  \text{if}~ r < R_{IJ} - D_{IJ}, \\[2ex]
\displaystyle \frac{1}{2} + \frac{1}{2} \cos \left( \frac{\pi(r-R_{IJ}+D_{IJ})}{2 D_{IJ}} \right) \qquad & \text{if}~ R_{IJ}-D_{IJ} \leq r \leq R_{IJ} + D_{IJ},   \\[2ex]
0 & \text{if}~ r > R_{IJ} + D_{IJ}.
\end{array}\right.
\end{equation*}
The repulsive potential $f_{\rm R}$ and the attractive potential $f_{\rm A}$ are given as
\begin{equation*}
f_{\rm R}(r;I,J) = A_{IJ} \exp(-\lambda_{IJ} r)  \qquad \text{and} \qquad f_{\rm A} (r;I,J) = -B_{IJ} \exp(-\mu_{IJ} r).
\end{equation*}
All material-specific parameters used in our numerical experiments are listed in the following tables.

\begin{table}[ht]
	\centering
	\caption{Two-index parameters for C-Si interactions \cite{tersoff94chemical}}
    \vskip 0.1cm
	\begin{tabular}{@{}p{1.5cm}p{2cm}p{2cm}p{2cm}p{2cm}p{2cm}p{2cm}@{}}
		\toprule
		\textbf{} & \textbf{$R$} & \textbf{$D$} & \textbf{$\lambda$} & \textbf{$A$} & \textbf{$\mu$} & \textbf{$B$} \\ 
		\midrule
		\multirow{1}{5cm}{C C} &1.95	&	0.15	&   3.4653  & 1544.8 & 2.3064 & 389.63 \\
		\midrule
		\multirow{1}{5cm}{Si Si} &2.85	&	0.15	&   2.4799 & 1830.8 & 1.7322 &  471.18 \\
		\midrule
		\multirow{1}{5cm}{C Si} & 2.35726	&0.15271	 &  2.9726 & 1681.731 & 2.0193 & 432.154 \\
		\bottomrule
	\end{tabular}
\end{table}

\begin{table}[ht]
	\centering
	\caption{Two-index parameters for C-B-N interactions \cite{kinaci2012thermal, sevik2011characterization}}
    \vskip 0.1cm
	\begin{tabular}{@{}p{1.5cm}p{2cm}p{2cm}p{2cm}p{2cm}p{2cm}p{2cm}@{}}
		\toprule
		\textbf{} & \textbf{$R$} & \textbf{$D$} & \textbf{$\lambda$} & \textbf{$A$} & \textbf{$\mu$} & \textbf{$B$} \\ 
		\midrule
		\multirow{1}{5cm}{C C} &1.95 & 0.15 & 3.4879 & 1393.6 & 2.2119 & 430.0 \\
		\midrule
		\multirow{1}{5cm}{C B} & 1.95 & 0.10 & 3.5279 & 1386.78 & 2.2054 & 339.06891 \\
		\midrule
		\multirow{1}{5cm}{C N} & 1.95 & 0.10 & 3.5279 & 1386.78  & 2.2054 & 387.575152 \\
        \midrule
		\multirow{1}{5cm}{B B} & 2.0 & 0.10 & 2.2372578 & 40.0520156 & 2.0774982 & 43.132016 \\
        \midrule
		\multirow{1}{5cm}{B N} & 1.95 & 0.05 & 3.568 & 1380.0 & 2.199 & 340.0 \\
        \midrule
		\multirow{1}{5cm}{N N} & 2.0 & 0.10 & 2.8293093 & 128.86866 & 2.6272721 & 138.77866 \\        
		\bottomrule
	\end{tabular}
\end{table}

\bibliography{bib}

\end{document}

%% file: notations.tex
\usepackage[bbgreekl]{mathbbol}
\usepackage{mathrsfs}
\usepackage{graphicx}
\usepackage{geometry}
\usepackage{amsmath}
\usepackage{amsfonts}
\usepackage{amssymb}
\usepackage{amsthm}
\usepackage{epstopdf}
\usepackage{xcolor}
\usepackage{bm}
\usepackage{float}
\usepackage{enumerate}
\usepackage[normalem]{ulem}
\usepackage{placeins}
\usepackage{algorithm}
\usepackage{algpseudocode}
\usepackage{empheq}
\usepackage{braket}
\usepackage{comment}
\usepackage{multirow}
\usepackage{array} 
\usepackage{booktabs} 
\usepackage{subfigure}
\usepackage{url}
\usepackage[normalem]{ulem}
\usepackage{cancel}
\usepackage{bbm}
\usepackage[colorlinks,linkcolor=blue,anchorcolor=green,citecolor=blue]{hyperref}

\renewcommand{\theequation}{\arabic{section}.\arabic{equation}}

\renewcommand{\thealgorithm}{\arabic{section}.\arabic{algorithm}}

\newcommand{\R}{\mathbb{R}}
\newcommand{\N}{\mathbb{N}}
\newcommand{\X}{\mathbb{X}}

\newcommand{\LL}{\mathbb{L}}

\newcommand{\Sset}{\mathbb{S}}

\newcommand{\A}{\mathbb{A}}

\newcommand{\config}{\bm{\sigma}}